\documentclass[preprint,12pt]{aastex}
\pdfoutput=1
\usepackage{natbib}
\usepackage{graphicx} 
\bibpunct{(}{)}{;}{a}{}{,}

\newcommand{\ee}[1]{\mbox{${} \times 10^{#1}$}}% scientific number format
\newcommand{\h}{\mbox{$^h$}}
\newcommand{\m}{\mbox{$^m$}}
\newcommand{\s}{\mbox{$^s$}}
\newcommand{\degree}{\mbox{$^{\circ}$}}

\newcommand{\kms}{\mbox{km s$^{-1}$}}% km/s

\newcommand{\lsun}{\mbox{L$_\odot$}}% Lsun
\newcommand{\msun}{\mbox{M$_\odot$}}% Msun
\newcommand{\form}{H$_2$CO}

\newcommand{\hcop}{HCO$^+$}

\newcommand{\dcop}{DCO$^+$}
\newcommand{\jj}[2]{\mbox{$J = #1\rightarrow#2$}}
\begin{document}

%%%%%%%%%%%%%%%%%% Title and Authors %%%%%%%%%%%%%%%%%%%%%%%%%%%%%%%%%%
\title {The Spitzer c2d Survey of Nearby Dense Cores XI : Infrared and
 Submillimeter Observations of CB130}

\author{
Hyo Jeong Kim\altaffilmark{1,2},
Neal J. Evans II\altaffilmark{1},
Michael M. Dunham\altaffilmark{1,3},
Jo-Hsin Chen\altaffilmark{1,4},
Jeong-Eun Lee\altaffilmark{5},
Tyler L. Bourke\altaffilmark{6},
Tracy L. Huard\altaffilmark{7},
Yancy L. Shirley\altaffilmark{8, 9},
Christopher De Vries\altaffilmark{10}
}

\altaffiltext{1}{Department of Astronomy, The University of Texas at
Austin, 1 University Station, C1400, Austin, Texas 78712--0259}

\altaffiltext{2}{E-mail: hyojeong@astro.as.utexas.edu}

\altaffiltext{3}{Department of Astronomy, Yale University, 260 Whitney Ave New Haven, CT 06511}

\altaffiltext{4}{Jet Propulsion Laboratory, 4800 Oak Grove Drive, Pasadena, CA 91109}

\altaffiltext{5}{Department of Astronomy and Space Science, Astrophysical
Research Center for the Structure and Evolution of the Cosmos,
Sejong University, Seoul 143-747}

\altaffiltext{6}{Harvard-Smithsonian Center for Astrophysics, 60
Garden Street, Cambridge, MA 02138}

\altaffiltext{7}{Department of Astronomy, The University of
Maryland, College Park, MD 20742-2421, USA}

\altaffiltext{8}{Steward Observatory University of Arizona 933
 N. Cherry Ave. Tucson, AZ 85721}

\altaffiltext{9}{Adjunct Astronomer at the National Radio Astronomy Observatory}

\altaffiltext{10}{California State University Stanislaus One University
 Circle Turlock, CA 95382 }

%%%%%%%%%%%%%%%%%%%% Abstract %%%%%%%%%%%%%%%%%%%%%%%%%%%%%
\begin{abstract}
We present new observations of the CB130 region, composed of three
separate cores. Using the \textit{Spitzer Space Telescope} we detected
a Class 0 and a Class II object in one of these, CB130-1. The observed
photometric data from \textit{Spitzer} and ground-based telescopes are
used to establish the physical parameters of the Class 0 object. SED
fitting with a radiative transfer model shows that the luminosity of
the Class 0 object is 0.14 $\--$ 0.16 \lsun, which is
a low luminosity for a protostellar object. In order to constrain
the chemical characteristics of the core having the low luminosity
object, we compare our molecular line observations to models of
lines including abundance variations. We tested both ad hoc step
function abundance models and a series of self-consistent chemical
evolution models. In the chemical evolution models, we consider a
continuous accretion model and an episodic accretion model to
explore how variable luminosity affects the chemistry. The step
function abundance models can match observed lines reasonably well.
The best fitting chemical evolution model requires episodic accretion
and the formation of CO$_2$ ice from CO ice during the low luminosity
periods. This process removes C from the gas phase, providing a much
improved fit to the observed gas-phase molecular lines and the CO$_2$
ice absorption feature. Based on the chemical model result, the low
luminosity of CB130-1 is explained better as a quiescent stage
between episodic accretion bursts rather than being at the first
hydrostatic core stage.
\end{abstract}

%\keywords{}

%%%%%%%%%%%%%%%%%%% Main text %%%%%%%%%%%%%%%%%%%%%%%%%%%%

\section{Introduction}\label{intro}
The \textit{Spitzer Space Telescope} Legacy Project, From Molecular
Cores to Planet Forming Disks (c2d, \citealt{2003PASP..115..965E})
has completed a survey of nearby star-forming regions. It covered
five clouds, containing 1024 objects classified as young stellar
Objects (YSOs) \citep{2009ApJS..181..321E}. When a dense core forms a
central protostar, it develops a luminosity source from the accretion
of infalling mass. The accretion luminosity is given as $L_{acc} = G
M_* \dot{M}_{acc} / R_*$, where $M_*$ is the mass of the protostar,
$\dot{M}_{acc}$ is the mass accretion rate onto the protostar, and
$R_*$ is the radius of the protostar. According to the standard model,
the mass accretion rate from spherical infall of the envelope is
$\dot{M}_{acc} \simeq 2 \times 10^{-6}$ \msun yr$^{-1}$ if the infall
occurs at the thermal sound speed at 10 K
\citep{1977ApJ...214..488S}. With a typical protostellar radius of 3
R$_{\odot}$, and a mass at the stellar/brown dwarf boundary of 0.08
\msun, the resulting $L_{acc}$ is 1.6 \lsun. Any luminosity from
contraction of the forming star will add to this accretion luminosity,
making it a minimum value in the standard model.

In contrast to the predictions, the c2d survey showed that 59\% of the
112 embedded protostars (Class 0/I) have luminosity lower than 1.6 \lsun
~\citep{2009ApJS..181..321E}, indicating that
either $M_*$ is very small, or $\dot{M}_{acc}$ is lower than expected,
or both. This luminosity problem was aggravated by the discovery of
Very Low Luminosity Objects (VeLLOs,
\citealt{2007prpl.conf...17D}). VeLLOs are defined as embedded
protostars with internal luminosity lower than 0.1 \lsun. The internal
luminosity of an embedded protostar is the luminosity of the protostar
and  disk, excluding luminosity from external heating. By using the
correlation between 70 $\micron$ flux and the internal luminosity of
protostars, \citet{2008ApJS..179..249D} identify 15 VeLLO candidates
 in the full c2d sample. Several VeLLOs including L1014-IRS
\citep{2004ApJS..154..396Y}, L1148-IRS \citep{2005AN....326..878K},
L1521F-IRS \citep{2006ApJ...649L..37B}, IRAM 04191-IRS
\citep{1999ApJ...513L..57A, 2006ApJ...651..945D}, L328-IRS
\citep{2009ApJ...693.1290L}, and L673-7 \citep{2010ApJ...721..995D},
have been studied in detail. Among those VeLLOs, IRAM 04191$+$1522 and
L673-7 drive strong molecular outflows \citep{1999ApJ...513L..57A,
 2006ApJ...651..945D, 2010ApJ...721..995D}. A low accretion
luminosity with a significant molecular outflow can imply higher
$\dot{M}_{acc}$ in the past \citep{2010ApJ...721..995D}.

One possible explanation for the sources with
low luminosities combined with  strong
molecular outflows is that the mass accretion is not a constant
process but rather episodic, and the VeLLOs are in quiescent phases
between the mass accretion bursts \citep{1990AJ.....99..869K,
2009ApJ...692..973E, 2010ApJ...710..470D}.
\citet{2010ApJ...710..470D} present a set of evolutionary models
describing collapse including episodic accretion. The luminosity
distribution of YSOs can be matched only when the model includes
episodic accretion.

While dust continuum emission can provide the physical structure and
evolutionary stage of a core, molecular line observations trace the
dynamics and chemistry of the core. The simple empirical models of
step function \citep{2003ApJ...583..789L} and drop function
\citep{2004A&A...416..603J} abundance profiles have been used to
approximate the abundance profiles. The chemical evolution model
during protostellar collapse developed by \citet{2004ApJ...617..360L}
has been tested for individual cores, like B335 \citep{2005ApJ...626..919E},
and L43 \citep{2009ApJ...705.1160C}. Using this model,
\citet{2007JKAS...40...85L} studied chemical evolution in VeLLOs.
The low luminosity sources can provide chemical laboratories to test the
astrochemistry of low luminosity environments. Because the chemical
equilibrium time can be longer than the duration of an accretion
burst, the abundance profile may provide a fossil record of previous
luminosity bursts.

This paper presents new observations of the CB130 region. With a detailed
analysis of \emph{Spitzer}, submm continuum, and molecular line data we will
reveal the physical properties of this source and use CB130 as a
laboratory to study the chemistry in low luminosity sources. In \S
\ref{CB130-1} we give a general introduction to CB130. The
observations are described in \S \ref{obs_sec}. \S \ref{result_sec}
presents images, photometry, and molecular line data. \S
\ref{radmodel_sec} presents dust continuum radiation transfer models
used to determine physical parameters. The line modeling with an
empirical step function is described in \S \ref{line_sec}. The
chemical evolution models with different luminosity evolution are
presented in \S \ref{chemical_evol_model}, and we summarize our
findings in \S \ref{summary}.

\section{CB130}\label{CB130-1}
\begin{figure}[t]
\includegraphics[width=0.9\columnwidth]{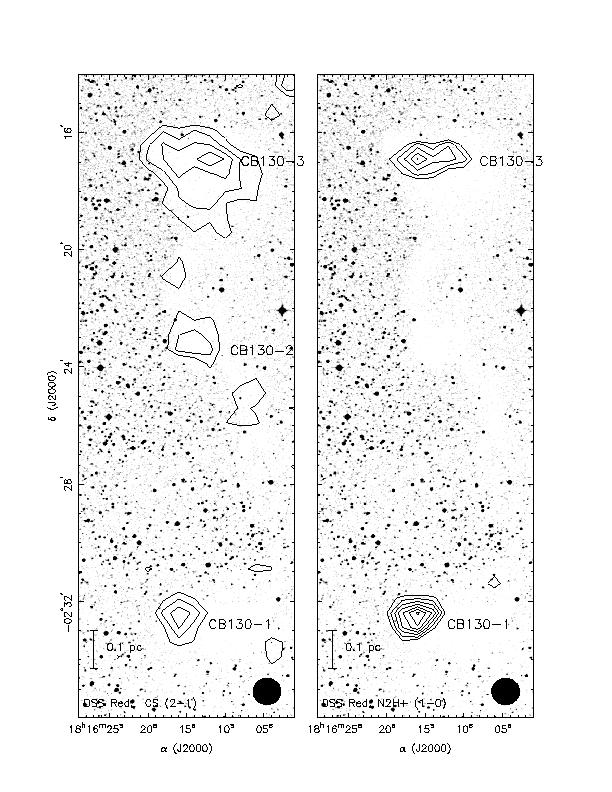}
\caption{\label{dss_img} The DSS R band image (gray scale)
with CS (2-1) (left) and N$_2$H$^+$ (1-0) (right) contours from the
FCRAO (De Vries et al., in preparation). The CS (2-1) contour levels
are 0.4, 0.6, 0.8 times the peak value of 0.32 K km s$^{-1}$. The
N$_2$H$^+$ (1-0) contour levels are 0.3, 0.4, 0.5, 0.6, 0.7, 0.8, 0.9
times of the peak value of 1.37 K km s$^{-1}$. N$_2$H$^+$ (1-0)
emission is not detected from the CB130-2 core. The beam size is shown
in the lower right corner of both panels.}
\end{figure}

CB130-1 first appears in the catalog of \citet{1988ApJS...68..257C},
which presents 248 small isolated molecular clouds.
\citet{1988ApJS...68..257C} called the core CB130. Later,
\citet{1999ApJS..123..233L} identified 3 cores near CB130 and named
them CB130-1, CB130-2, and CB130-3, respectively, from south to
north. Fig.~\ref{dss_img} shows the R-band image from the Digital Sky
Survey (DSS), with CS (left) and N$_2$H$^+$ (right) molecular line
contours (De Vries et al., in prep, see \S \ref{obs_sec}) overlaid. The
three cores are clearly identified as separate dark cores in the
region. \citet{2007AJ....133.1560W} presented Submillimeter High
Angular Resolution Camera II (SHARC-II) 350 $\micron$
and 450 $\micron$ detections of CB130-1. \citet{2008ApJS..179..249D}
identified an embedded source in CB130-1 with L$_{int} \sim$ 0.07 \lsun~
based on a correlation between 70 $\micron$ flux and the internal
source luminosity. However, this result is an estimate, and is
only good to within a factor of $\sim 2$. Recently,
\citet{2010ApJS..188..139L} presented a survey of low mass star
forming cores in 32 Bok globules, including CB130-1. They found a faint
near infrared (NIR) source and a very red star in CB130-1, and they
identified these as Class 0 and Class II objects, respectively.

CB 130 is located in the Aquila Rift cloud, whose galactic
coordinates are $20^{\circ} <l< 40^{\circ}$ and $-6^{\circ} <b<
+14^{\circ}$. The local
standard of rest velocity of the Aquila Rift cloud is 8 km/s
\citep{1985ApJ...297..751D}, which is similar to the velocity of
CB130-1 (see \S \ref{line_sec}). \citet{2003A&A...405..585S}
determined the distance to the Aquila Rift, using extinction methods,
to extend between 225 and 310 pc, and the central region at 270 pc. We
thus adopt a distance to CB130-1 of 270 pc but
note the substantial uncertainty.

\section{Observations}\label{obs_sec}
We obtained observational data from various telescopes. All three
instruments mounted on \emph{Spitzer} observed CB130-1 in the c2d
program. The Infrared Array Camera (IRAC;
\citealt{2004ApJS..154...10F}), the Infrared Spectrograph (IRS;
\citealt{2004ApJS..154...18H}), and the Multiband Imaging Photometer
for SIRTF (MIPS; \citealt{2004ApJS..154...25R}). The IRAC images were
obtained on September 2nd 2004 and September 3rd
2004 (Program ID [PID] 139; c2d \citealt{2003PASP..115..965E};  AOR
keys 0005145856 and 0005146368), the IRS spectra were obtained on
April 26th 2006 (PID 179; PI: N. J. Evans; AOR key 0015921153), and
the MIPS 24 and 70 $\micron$~ data were obtained on September 25th 2004
(PID 139, AOR key 0009412608, 0009421824).

The IRAC and MIPS images from c2d were reduced by the
\emph{Spitzer} Science Center (SSC) using the standard pipeline
version S13 to produce Basic Calibrated Data (BCD) images. The BCD
images are obtained by subtracting the dark and bias levels, and
performing flat fielding and sky subtraction. A full explanation of
this extra processing data can be found in the documentation of the
final delivery of data from the c2d legacy project
\citep{2007..c2d..IRAC}.

The IRS data were reduced following the IRS pipeline version
S14.0.0, which produces BCD files. IRS BCD files are two dimensional
spectra, which have been processed including saturation flagging,
dark-current subtraction, linearity correction, cosmic ray
correction, ramp integration, droop correction, stray light removal
or crosstalk correction, and flat-field correction. Detailed
explanations about the data delivery can be found in the c2d
Spectroscopy Explanatory Supplement \citep{2006..c2d..IRS}.

\emph{Spitzer} IRAC and MIPS 24 $\micron$ images of CB130-1 were
also obtained in the \emph{Spitzer} GO-2 program (cores2deeper;
PID 20386; PI: P. C. Myers), which obtained deeper images of a sample
of cores from the c2d program. IRAC 4 band images were obtained on
September 16th 2005 (AOR key 0014606848) and on September 19th 2005
(AOR key 0014607104). MIPS 24 $\micron$~ image was obtained on October
5th 2005 (AOR key 0014615040). Additional \textit{Spitzer} MIPS images
at 24 $\micron$~ and 70 $\micron$~ were obtained along with 160
$\micron$~ image on May 18th 2007 (PID 30384; PI: T. L. Bourke; AOR
key 0018159616). IRAC images were reduced in the same way as the c2d
data. MIPS images from PID 20386 and PID 30384 were reduced using the
MIPS Data Analysis Tool (DAT; \citet{2005PASP..117..503G}). The data
reduction process is described in \citet{2007ApJ...665..466S}.

Submm Common User Bolometer Array 
(SCUBA)\footnote{This research used the facilities of the Canadian
 Astronomy Data Centre operated by the National Research Council of
 Canada with the support of the Canadian Space Agency.} observations
at 850 \micron, taken at the
James Clerk Maxwell Telescope (JCMT), were obtained
from the CADC archive.  The SCUBA
jiggle maps were analyzed using the standard SCUBA User Reduction
 Facility (SURF) reduction package and aperture photometry was
 calibrated using Uranus observations obtained from the same night
 (see \citealt{2000ApJS..131..249S} for the flux calibration
procedure). The beam size at 850 $\micron$ is $15''$.

Near-infrared observations of CB130-1 were obtained during 24-25
March 2005 using the Infrared Sideport Imager (ISPI) on the 4-meter
Blanco telescope at Cerro Tololo Interamerican Observatory (CTIO).
Deep observations at J and H were obtained with 60-second and
30-second exposures, respectively. Deep Ks observations were
obtained with 2 coadded 10-second exposures.  The total exposure
times achieved for these deep observations were 9, 76, and 17
minutes at J, H, and Ks, respectively. To obtain photometry of most
moderately bright stars, which were saturated in these deep
observations, shallower 3-second exposures were obtained. The
individual images were reduced using IRAF and IDL procedures
following the standard method, as described in Huard et al. (2006),
to construct the final stacked images at J, H, and Ks. Sources were
identified in these final images and $1\farcs2$-radius aperture
photometry derived using PhotVis, version 1.09
\citep{2004ApJS..154..374G}, an IDL GUI-based aperture photometry
tool that uses standard DAOPHOT packages
\citep{1993ASPC...52..246L}. Finally, the photometry was calibrated
by comparison with published photometry of sources listed in the Two
Micron All Sky Survey (2MASS) catalog. The average seeing during the
J observations was $\sim$ $1\farcs3$, and $\sim$ $1\farcs1$ during
the H and Ks observations.

The molecular line spectral data were obtained from several
telescopes (see Table ~\ref{line_obs_cb130-1} \--
~\ref{line_obs_cb130-3}). The majority of the lines were observed at
the Caltech Submillimeter Observatory (CSO) \footnote{The Caltech
 Submillimeter Observatory is  supported by the NSF.} from 2005 to
2009. For the data obtained between 2005 \-- 2007, we used an
acousto-optic spectrometer (AOS), with 1024 channels and 50 MHz
bandwidth. For the 2008 \-- 2009 data, we used a fast fourier
transform spectrometer (FFTS) with 8192 channels and 500 MHz
bandwidth. We used the 230 GHz heterodyne receiver for all the lines
except H$_2$D$^+ ~(J_{K_{-1} K_{+1}} = 1_{1,0} \rightarrow 1_{1,1})$,
for which we used the 345 GHz receiver. The position switched scans
were taken when the optical depth at 225 GHz was between 0.1 and
0.2. The calibration uncertainty is about 10\%. N$_2 $H$^+ (J= 1
\rightarrow 0)$ and CS $(J=2 \rightarrow 1)$ maps were obtained at the
Five College Radio Astronomy Observatory (FCRAO; De Vries et al., in
prep) and N$_2 $H$^+ ~( J= 1 \rightarrow 0)$ data were obtained from
the Arizona Radio Observatory (ARO) on March 2007. We reduced the
molecular line data from the CSO, FCRAO, and ARO 12 m telescope using
CLASS \footnote{http://www.iram.fr/IRAMFR/GILDAS}.

\section{Results}\label{result_sec}
\subsection{The Images and Photometry}
\begin{figure}[c]
\includegraphics[width=0.9\columnwidth]{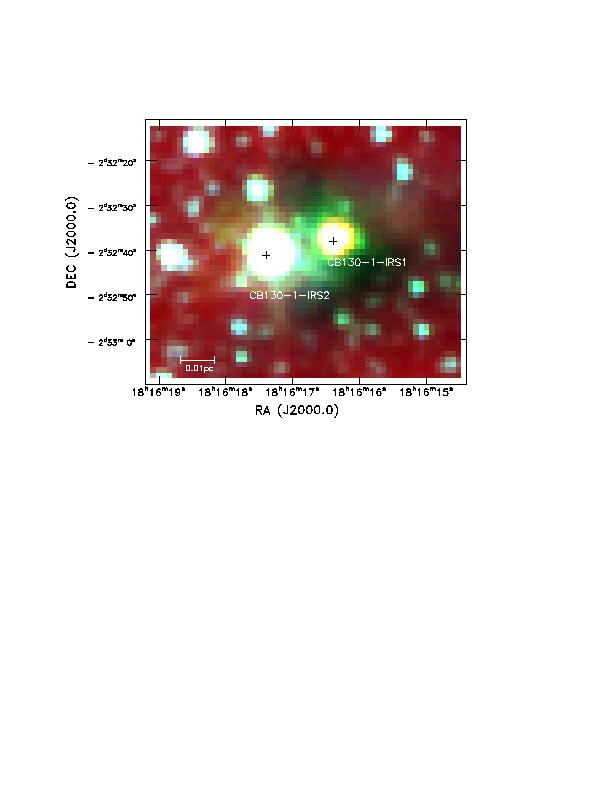}
\caption{\label{spitzer_image} Three color image of CB130-1 using the
cores2deeper IRAC image with IRAC 3.6 $\micron$, 4.5 $\micron$, and
8.0 $\micron$ as blue, green and red, respectively. CB130-1-IRS1
and IRS2 are labeled and marked with black plus signs.}
\end{figure}

A three-color image of CB130-1 using cores2deeper IRAC data is
presented in Fig.~\ref{spitzer_image}, with 3.6 $\micron$ as blue,
4.5 $\micron$ as green, and 8.0 $\micron$ as red. Two YSOs are
identified in these data based on their positions in various
color-color \& color-magnitude diagrams
\citep{2007ApJ...663.1149H}. These two YSOs are $15 ''$ apart, which
corresponds to 4100 AU at a distance of 270 pc
\citep{2003A&A...405..585S}. We name the western source
($\alpha$ = 18$\h$ 16$\m$ $16.4 \s$, $\delta$ = $-02^{\circ} ~32'
~38\farcs 0$ [J2000]) CB130-1-IRS1 and the eastern source ($\alpha$
= 18$\h$ 16$\m$ $17.4 \s$, $\delta$ = $-02^{\circ} ~32' ~41\farcs 1$
[J2000]) CB130-1-IRS2 and label these sources in Fig.~\ref{spitzer_image}

\begin{figure}
\includegraphics[width=0.9\columnwidth]{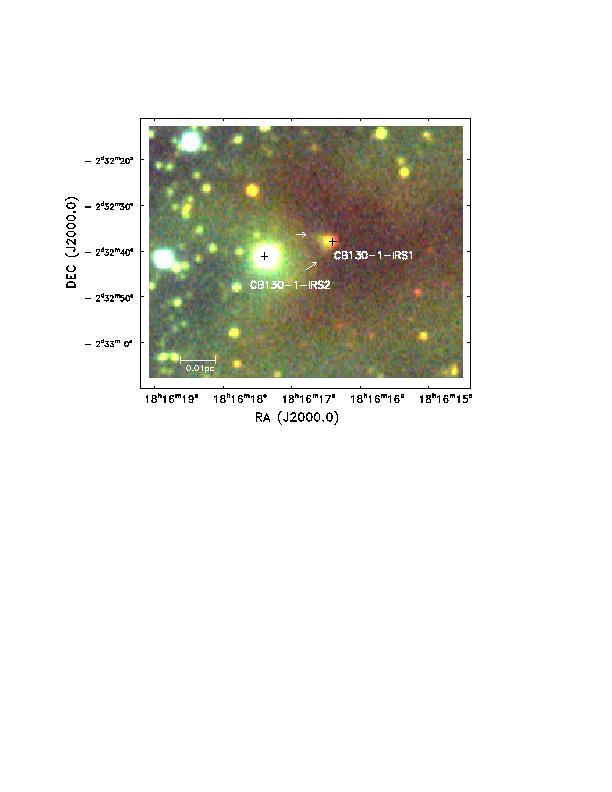}
\caption{\label{IR_map} The three color image composed of near IR
band image, with J as blue, H as green, and K as red. The left plus
sign indicates CB130-1-IRS2, and the right plus sign is the
position of CB130-1-IRS1. The light at the position of CB130-1-IRS1
is mostly scattered light from CB130-1-IRS1. The white arrows
indicate the cone shape nebulosity from CB130-1-IRS1 extending toward
CB130-1-IRS2.}
\end{figure}

Fig.~\ref{IR_map} shows a three color image of CB130-1 using the NIR
data  with J as blue, H as green and Ks as red. Both YSOs are detected
and labeled in the figure. Cone shaped nebulosity extends from
CB130-1-IRS1 toward CB130-1-IRS2, as marked by arrows in the
figure. Similar nebulosity is also observed in L1014
\citep{2006ApJ...640..391H} and is indirect evidence of an
outflow. However, we mapped the CB130-1 region with CO (\jj{2}1) at
the CSO and found no significant evidence of outflowing gas.

\begin{figure}[t]
\includegraphics[scale=0.7]{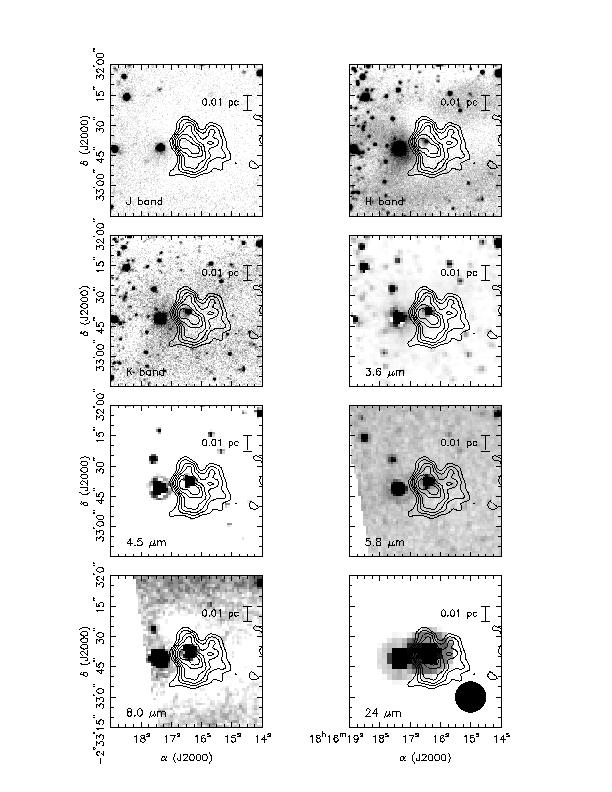}
\caption{\label{850_cont} The photometry images from the different
wavelengths. The J, H, K band images are from ISPI, the 3.6
$\micron$, 4.5 $\micron$, 5.8 $\micron$, and  8.0 $\micron$ images
are from IRAC, and 24 $\micron$ image is from MIPS. We plot the
contours taken from the 850 $\micron$ SCUBA. The beam size of SCUBA at
850 $\micron$ is shown in the bottom right of the 24 $\micron$
panel. The contour levels are 0.5, 0.6, 0.7, 0.8, 0.9 times of the
peak value (248 mJy/beam). }
\end{figure}

Fig.~\ref{850_cont} shows 850 $\micron$ contours overlaid on the 1.25
$\--$ 24 \micron~ NIR and \emph{Spitzer} images. The 850
$\micron$ data probes the envelope since dust emission from the
envelope becomes optically thin at long wavelengths. The 850
$\micron$ contours peak at CB130-1-IRS1 and avoid CB130-1-IRS2.
CB130-I-IRS2 is brighter at all wavelengths up to 8 \micron, but
CB130-1-IRS1 is stronger at 24 \micron.
These facts support the conclusion by \citet{2010ApJS..188..139L}
that CB130-1-IRS1 is younger and more embedded than CB130-1-IRS2.

\begin{figure}[t]
\includegraphics[width=0.9\columnwidth]{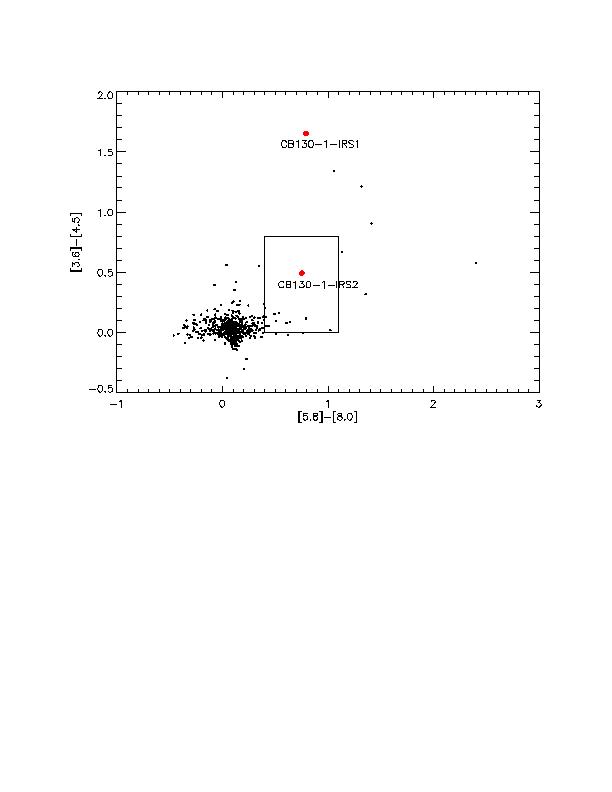}
\caption{\label{cc-diagram} A color-color diagram for the 4 IRAC
bands. The notation [3.6]$-$[4.5] denotes the magnitude difference,
or color, between 3.6 and 4.5 \micron. The box indicates the
location of Class II sources \citep{2004ApJS..154..363A}. The two
YSOs in CB130-1 are clearly redder than background stars. CB130-1-IRS2
is clearly in the box where Class II sources are found, whereas IRS1
is not. The black dots are classified as stars (those with colors near
zero) or galaxies (those in the Class II or Class I regions).
}
\end{figure}

A color-color diagram constructed from the IRAC photometry for all
sources detected in the \emph{Spitzer} images of the CB130 region
is plotted in Fig.~\ref{cc-diagram}. Sources inside the
box are generally Class II sources \citep{2004ApJS..154..363A}. The
two sources in CB130-1 are clearly redder than background stars. The
color-color diagram shows that CB130-1-IRS1 and CB130-1-IRS2 are in
different phases of evolution. CB130-1-IRS2 is clearly in the region
of Class II sources. The sources for which either the [3.6]-[4.5] color is
greater than 0.8 mag or the [5.8]-[8.0] color is greater than 1.1 mag
are all faint
and classified as galaxies except for CB130-1-IRS1
\citep{2007ApJ...663.1149H}. Since no YSO candidates in CB130-2 and
CB130-3 were detected with CTIO and \textit{Spitzer}, we will
concentrate on CB130-1 for the remainder of the paper.

\subsection{CB130-1-IRS1}\label{result_CB130-1-IRS1}
\begin{figure}[t]
\includegraphics[width=0.9\columnwidth]{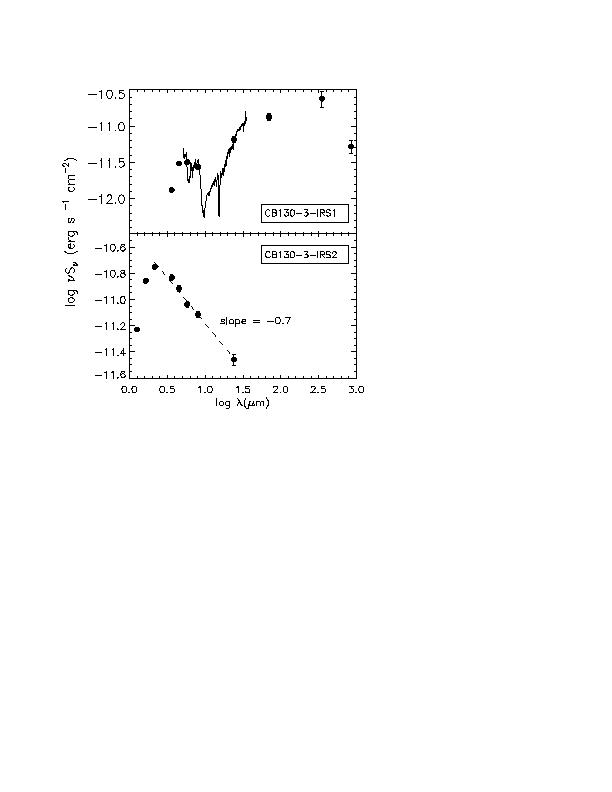}
\caption{\label{SED-observed}The SED of CB130-1-IRS1 (upper panel)
and CB130-1-IRS2 (lower panel). The photometry data from ISPI, IRAC,
MIPS, SHARC-II and SCUBA are plotted as filled circles with error
bars. The IRS spectrum of CB130-1-IRS1 is plotted as a solid
line. CB130-1-IRS1 has a generally rising infrared SED, and peaks at
350 $\micron$. CB130-1-IRS2 is not detected longward of 24
$\micron$. The dashed line in the lower panel is the slope of the SED
from 2.17 $\micron$ to 24$\micron$. The slope is $-0.69$, which
indicates CB130-1-IRS2 is a Class II object.}
\end{figure}

The photometric data for CB130-1-IRS1 are given in
Table~\ref{table-IRS1}, which lists wavelength, flux density,
uncertainty in flux density, aperture diameter, and telescope. At 3.6
$\--$ 24 $\micron$, we list the  flux averaged between the c2d and
cores2deeper images, weighted by the inverse of the uncertainties.
The SED is plotted in the upper panel of Fig.~\ref{SED-observed},
including the IRS spectrum, which clearly contains 10 $\micron$
silicate and 15.2 $\micron$ CO$_2$ absorption features. Using the 15.2
$\micron$ CO$_2$ absorption feature, we can derive the CO$_2$
abundance toward CB130-1-IRS1. The continuum for the spectrum is
constructed by fitting a first order polynomial to the ranges 12 $\--$
14.7 and 16.2 $\--$ 18 $\micron$. After subtracting the continuum, we
calculated the CO$_2$ column density from $N$(CO$_2$)$ = \int \tau
(\nu ) d\nu / A$, where $\tau (\nu )$ is an optical depth, $\nu$ is
the wave number (cm$^{-1}$), and $A = 1.1 \times 10^{-17}$ is the band
strength \citep{1995A&A...296..810G}. We find $\int \tau (\nu ) d\nu =
28 \pm 0.4$ cm$^{-1}$ and $N$(CO$_2$)$ = (2.6 \pm 0.05) \times
10^{18}$ cm$^{-2}$. We calculated the H$_2$ column density using 850
$\micron$ data. Using equation (6) in \citet{2003ApJ...583..789L}, we
have $N$(H$_2$)$= 4.5 \times 10^{22}$ cm$^{-2}$. From $N$(CO$_2$) and
$N$(H$_2$), the abundance of CO$_2$ ice is $5.8 \times 10^{-5}$, which
requires a substantial fraction of the carbon.

Based on the observed SED, we can calculate standard observational
signatures including $L_{bol}$, $T_{bol}$, and
$L_{bol}/L_{(>350\micron)}$. We calculate $L_{bol}$ by
integrating the flux over all wavelengths using the average of the
results obtained from midpoint and prismoidal integration methods.
$T_{bol}$ is the temperature of a blackbody having
the same flux-weighted mean frequency as the source and is calculated
following \citet{1993ApJ...413L..47M}. $L_{(>350\micron)}$ is the
luminosity emitted at $ \lambda >350 ~\micron $. CB130-1-IRS1 has
$L_{bol} = 0.23 \pm 0.03$ \lsun, $T_{bol} = 59 \pm 1.6$ K, and
$L_{bol}/L_{(>350\micron)} = 9.1$. $L_{bol}/ L_{(>350\micron)}$ is used
 to distinguish Class 0 from Class I sources. The pre-protostellar
core/Class 0 transition occurs around $L_{bol} / L_{(>350\micron)}
\sim 35$ and the Class 0/I transition occurs at $L_{bol}/
L_{(>350\micron)} \sim 175$ \citep{2005ApJ...627..293Y}. Since
$L_{bol}/L_{(>350\micron)}$ is less than 35, we would guess that
CB130-1 is in the pre-protostellar core stage from these criteria, but
CB130 clearly contains embedded sources.
Another criterion to define a Class 0 object is the
bolometric temperature of the source. Class 0 and I objects are
divided at $T_{bol} = 70$ K \citep{1995ApJ...445..377C}; thus
CB130-1-IRS1 is classified as a Class 0 source based on $T_{bol}$.

The bolometric luminosity includes the luminosities of the embedded
object, the disk and the ISRF, which can add several tenths of a
\lsun~ to $L_{bol}$~ \citep{2001ApJ...557..193E}. The luminosity of
the internal source can be estimated in two ways. First, assuming that
all emission at $\lambda \ge 100 $ \micron\ arises from external
heating, $L_{int}$ can be calculated by integrating the flux at
$\lambda \le 100 $ \micron, giving $L_{int} \sim $ 0.056
\lsun. Alternatively, we can estimate $L_{int}$ by applying the
relation between the flux at 70 $\micron$, normalized to 140 pc, and
$L_{int}$ found by \citet{2008ApJS..179..249D}, giving $L_{int} \sim
0.07 ~\lsun$. These two estimates indicate that CB130-1-IRS1 is in any
case a low luminosity object and may be a VeLLO. However, they are
only rough estimates; radiative transfer models of CB130-1 are
required to reliably determine the physical properties of the internal
source (\S \ref{radmodel_sec}).

Assuming that the core is optically thin at submillimeter
wavelengths, the envelope mass can be estimated from $M=
\frac{S_{\nu} D^2}{B_\nu (T_d ) \kappa_\nu}$, where $S_{\nu}$ is the
flux density at 850 \micron, D is the distance to the object,
$B_\nu (T_d )$ is the  black-body function, and $\kappa_\nu$ is the
opacity of gas and dust at 850 $\micron$
\citep{2003ApJS..145..111Y}. Pontoppidan's dust model (Pontoppidan et
al., 2010, in preparation) gives $\kappa_\nu = 1.02 \times 10^{-2}$
cm$^{-1}$ g $^{-1}$, and $\kappa_\nu = 1.8 \times 10^{-2}$ cm$^{-1}$ g
$^{-1}$ for OH5 dust \citep{1994A&A...291..943O}. If we assume the
core is isothermal at $T_d = 10$ K, the estimated envelope mass within
a 120\arcsec\ diameter aperture (radius of 16,200 AU) is 3.5~\msun\
with the Pontoppidan model, and 2.0~\msun\ with the OH5 model. These
estimates will be refined later.

\subsection{CB130-1-IRS2}
The SED of CB130-1-IRS2 is plotted in the lower panel of
Fig.~\ref{SED-observed}. CB130-1-IRS2 is not detected at $\lambda > 24
\micron$. The slope of the SED of CB130-1-IRS2 from 2.17 $\micron$ to
24 $\micron$ is $-0.7$, which indicates that CB130-1-IRS2 is a Class
II object \citep{1994ApJ...434..614G}. We can estimate the bolometric
luminosity of the source by integrating the SED. The bolometric
luminosity of CB130-1-IRS2 is $L_{bol} = 0.071 \pm 0.003$ \lsun\ and
the bolometric temperature is $T_{bol} = 1150 \pm 15$ K. These values
of $L_{bol}$ and $T_{bol}$ are not corrected for extinction along the
line of sight, and are thus lower limits to the intrinsic values.

\subsection{The Molecular Line Data}
\begin{figure}[t]
\includegraphics[width=0.9\columnwidth]{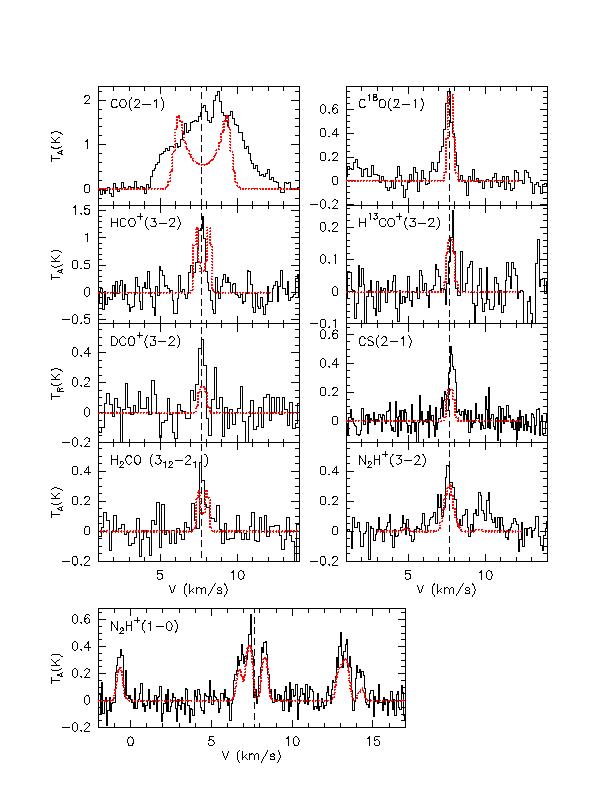}
\caption{\label{Line:model} The observed molecular lines of CB130-1
and line modeling results for step function abundances
(see \S \ref{line_sec}). The observed data are plotted as solid
lines. The vertical dashed lines are line centers determined by
averaging the LSR velocity of optically thin lines. The modeled data
are plotted as dotted lines. The modeled C$^{18}$O, H$^{13}$CO,
N$_2$H$^+$, and CS lines show a good fit with the observed lines,
while CO and HCO$^+$ lines show a poor fit with the observed lines.
CO is confused by the wing component; modeled lines of HCO$^+$ and
\form\ show self-absorption features.}
\end{figure}

\begin{figure}[t]
\includegraphics[width=0.9\columnwidth]{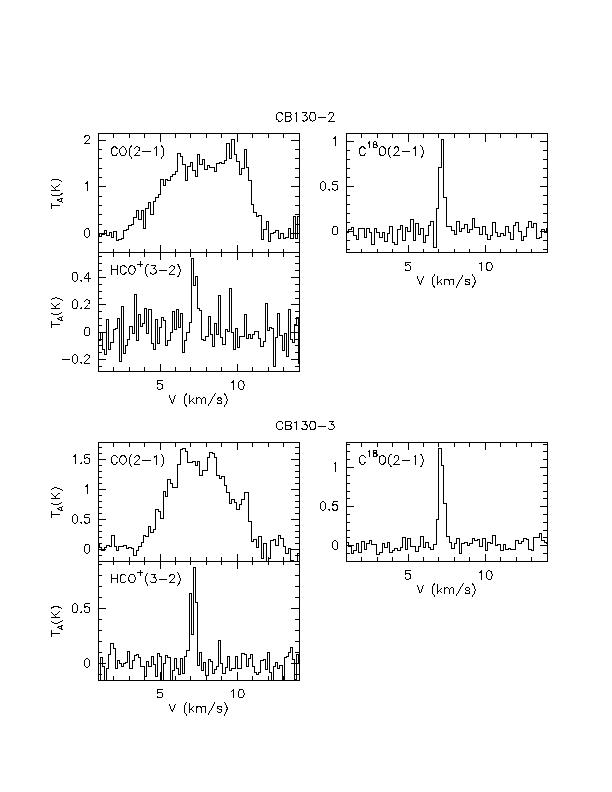}
\caption{\label{line_obs_cb130-23} The observed molecular lines of
CB130-2, and CB130-3. The parameters and descriptions are in
Table~\ref{line_obs_cb130-2} and Table~\ref{line_obs_cb130-3}. }
\end{figure}

Molecular line observations can trace the dynamics in the core,
such as infall, rotation, and outflow. The observed molecular lines
for CB130-1, CB130-2, CB130-3 are summarized in Table
\ref{line_obs_cb130-1}, Table \ref{line_obs_cb130-2}, and Table
\ref{line_obs_cb130-3} respectively. The detected molecular lines
are plotted in Fig.~\ref{Line:model} as solid lines. Toward CB130-2,
the detected molecular lines are plotted in the upper three panels in
Fig.~\ref{line_obs_cb130-23}. Toward CB130-3, the detected lines are
plotted in the lower three panels in Fig.~\ref{line_obs_cb130-23}.

In the next section, we will focus on CB130-1-IRS1, since the
source is a deeply embedded Class 0 object. We will find the
internal luminosity and chemical properties of the source.

\section{Radiative Transfer Model}\label{radmodel_sec}

As discussed above, a detailed radiative transfer model is required to
accurately determine the internal luminosity of CB130-1-IRS1. The
internal luminosity will determine whether the source
is a VeLLO or not.
This section describes one-dimensional and two-dimensional models of
this source. These models trace the radiation from a source in a dusty
region, including scattering, absorption, and re-emission by the dust
and generate model SEDs that can be compared to the observations.

The one-dimensional models provide one estimate of the internal
luminosity and a spherical model that is needed for analyzing the
molecular lines. The two-dimensional models provide an alternative
estimate of the internal luminosity and a better fit to the SED, at
the expense of having more free parameters.

\subsection{One-dimensional radiative transfer models}\label{DUSTY_sec}

We modeled CB130-1-IRS1 with the one-dimensional radiative
transfer code, DUSTY \citep{1999astro.ph.10475I}. DUSTY calculates
the dust temperature profile and SED of a protostar embedded in a
dense core. The parameters of the model divide into those associated with
the envelope, or core, and those associated with the internal source.

The envelope is characterized by the outer radius, inner radius,
density distribution, dust properties, and external radiation field.
The outer radius of the envelope is not well constrained by the
radiative transfer modeling as long as it is large enough to include
all the emission.  We fix the outer radius at 35000 AU,
based on the angular size of CB130-1 ($3\farcm0 \times 1\farcm3$,
\citealt{1999ApJS..123..233L}). The inner radius ($R_{in}$) will be
a free parameter.  We assume a power-law radial density distribution
for the envelope ($n = n_0(r/r_0)^{-\alpha}$), where $n_0$ is
the density at a fiducial radius, $r_0$. The value of $n_0$ was
constrained to match the long-wavelength emission, and
$\alpha = 1.8 \pm 0.2$ was constrained to match the radial intensity
profile. See \citet{2002ApJ...575..337S} for an explanation of this
method.

We use the Pontoppidan dust opacity model (Pontoppidan et al., 2010,
in preparation). The envelope is heated by both the embedded source
and the ISRF described by \citet{2001ApJ...557..193E}. The ISRF can
vary in strength and is modified by extinction by a surrounding low
density envelope. To constrain the ISRF, we used an initial model of
the envelope and modeled the CO ($J=2 \rightarrow 1$) line, since the CO
($J=2 \rightarrow 1$) line is sensitive to the UV part of the ISRF, and
thus the extinction. The best fit value for the UV part of ISRF is
$G_0 = 4.5 \times 10^{-3}$, corresponding to $A_{V} = 3$. We will
discuss this in \S \ref{line_sec}. With these dust properties, the
density normalization ($n_0$) yields the total mass of the model
envelope within 35,000 AU to be 5.3~\msun. This mass depends only
slightly on the best-fitting parameters for luminosity and inner
radius, as long as they are close to the best-fit values. Within
16,200 AU (the size used to estimate the observed mass in section
\ref{result_CB130-1-IRS1}), the model contains a mass of 2.2 \msun,
slightly less than the isothermal calculation within that radius.

\begin{figure}[t]
\includegraphics[width=0.89\columnwidth]{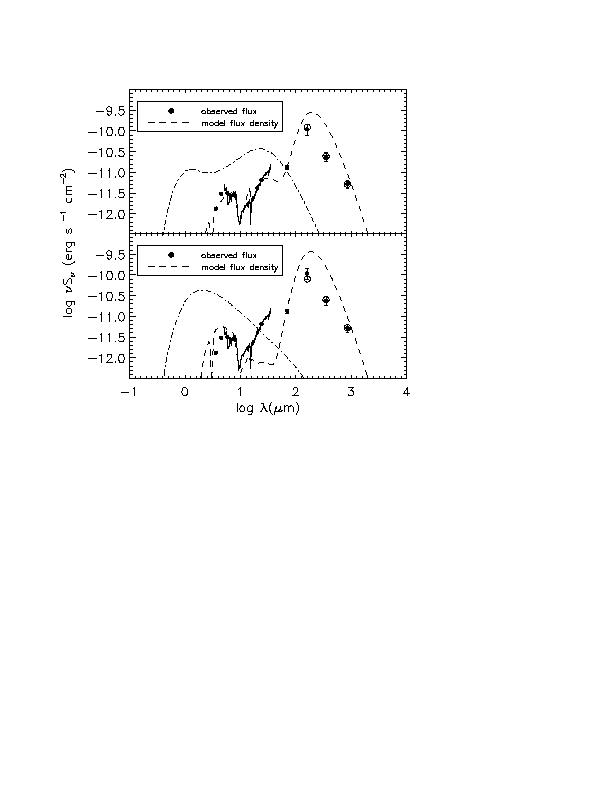}
\caption{\label{dusty:model} The photometry data and the best fit
1-D radiative transfer model result for CB130-1-IRS1. The photometry
data are plotted as filled circles. The data from the IRS is plotted
as a solid line. The dash-dot line is the input SED from the
internal source. The final SED is plotted as the dashed line. The
aperture size convolved SED is marked with open circles at
wavelengths longer than 100 $\micron$. The model parameters are
T$_s$ = 3000 K, $L_{int} = 0.16$ \lsun, $R_{in} = 350$ AU. The
lower panel indicates the radiative transfer modeling result with a
passive disk model. With a temperature distribution of a passive
disk model, we cannot fit the 24 $\micron$ data. }
\end{figure}

The internal source is characterized by an internal luminosity
($L_{int}$) and a stellar temperature ($T_s$). The luminosity of the
disk is included in the model, but the disk luminosity is
not easily separable from the stellar luminosity, so we treat the sum
as the free parameter: $L_{int} = L_s + L_{disk}$. The emission from
the disk is averaged over all inclination angles and added to the
input spectrum \citep{1994ApJ...420..326B}. The 24 $\micron$
photometry data constrains the disk parameters. The surface density
profile of the model disk follows $\Sigma (r) \propto r^{-p}$ with
$p=1.5$. We choose a disk temperature distribution following $T(r)
\propto r^{-q}$. When the disk is a flat passive disk, $q = 0.75$. For
the flat passive disk, all the luminosity comes from the re-radiation
from the central star. A flat passive disk produces less 24 $\micron$
emission than observed (lower panel of Fig.~\ref{dusty:model}). We
varied the index of the temperature distribution until it fitted the
24 $\micron$ observed data. The best fit power law index of the disk
temperature distribution is $q=0.4$, which corresponds to a flared disk
\citep{1997ApJ...490..368C}. The disk temperature distribution is also
less steep than the flat disk case if the disk has an intrinsic
luminosity source, such as accretion and back warming from the
envelope in addition to radiation from the central star.

DUSTY calculates the radiative transfer from the internal source and the
external radiation field. Then, the OBSSPHere
\citep{2002ApJ...575..337S} program, which convolves the model with a
beam, is applied to the DUSTY result to convert models into observed
flux densities and model radial profiles at wavelength longer than 100
\micron. The model parameters are constrained by the observed SED and
the observed radial intensity profile at 850 \micron.

We tested a grid of models to obtain the best-fit free parameters.
After the constraints described above, the remaining free parameters
are $T_{s}$, $L_{int}$, and $R_{in}$. We calculated the reduced
$\chi^2$ to quantify the quality of the fit. The data used to
constrain the model covers 3.5 $\micron  \leq \lambda \leq$ 850
$\micron$, with a total of nine points of photometry data. We do not
include H band photometry data, because H band flux comes mainly from
scattered photons in the envelope, and DUSTY modeling does not
properly treat scattering at short wavelengths. Also, we focus on the
photometric data, and check the IRS spectrum only for consistency. We
have 9 data points and 3 free parameters, so the degree of freedom of
reduced $\chi^2$ is 6.

\begin{figure}[t]
\includegraphics[width=0.9\columnwidth]{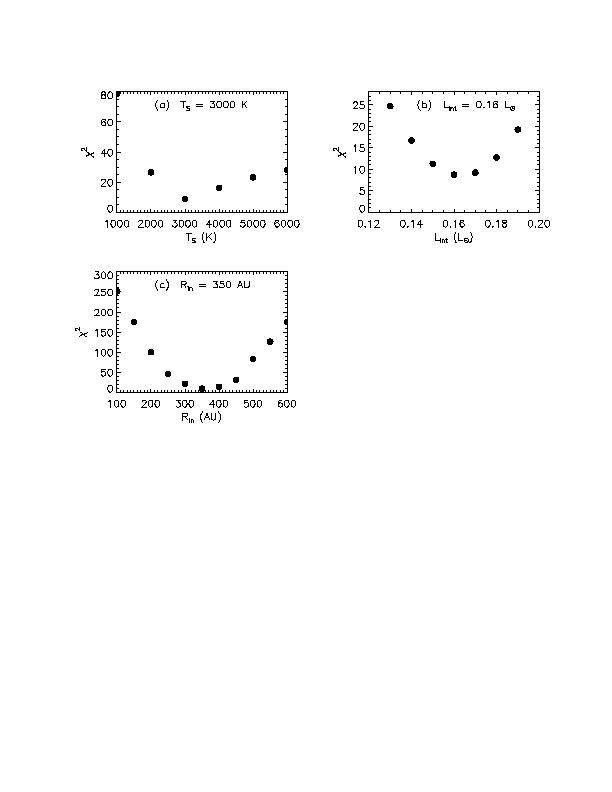}
\caption{\label{chisq:dusty} The $\chi^2$ plots for a grid  of
one-dimensional radiative transfer models. Panel (a) is
$\chi^2$ versus $T_s$, with $L_{int}$ and $R_{int}$ held fixed at 0.16
\lsun ~and 350 AU, respectively. The smallest $\chi^2$ is at $T_s = 3000$ K. Panel (b) is
$\chi^2$ versus $L_{int}$, $T_{s}$ and $R_{in}$ held fixed at 3000 K
and 350 AU. The smallest $\chi^2$ is at $L_{int} = 0.16$ \lsun. Panel
(c) is $\chi^2$ versus $R_{in}$, with $T_{s}$ and $L_{int}$ held fixed
at 3000 K, and 0.16 \lsun. The smallest $\chi^2$ is at $R_{in} = 350$ AU. }
\end{figure}

The $\chi^2$ plots are in Fig.~\ref{chisq:dusty}. When we plot
Fig.~\ref{chisq:dusty}, we hold fixed the other two parameters at the
values with the smallest $\chi^2$ values. We plot in panel (a) of
Fig.~\ref{chisq:dusty} varying  $T_s$ from 1000K to 7000 K.
The value of $T_s$ does not critically affect the modeled SED, since
all the stellar photons are reprocessed by dust close to the star.
The lowest $\chi^2$ is found at $T_{s}=3000 \pm 500 $ K. Panel (b) is
$\chi^2$ obtained by varying $L_{int} = L_{s} + L_{disk}$.  The best
fit is found at $L_{int}=0.16 \pm 0.005$ \lsun. The $L_{int}$ is a
little higher than the VeLLO criterion, but much lower than the
accretion luminosity calculated from the mass accretion rate in
\citet{1977ApJ...214..488S}'s model. Panel (c) is the $\chi^2$ plot
varying $R_{in}$.  The inner radius of the envelope controls the
optical depth. When the optical depth increases, the flux density
decreases at short wavelengths because dust absorbs energy at short
wavelengths and re-emits at long wavelengths. The best fit inner
radius is $R_{in} = 350 \pm 25$ AU, which gives $\tau = 0.22$ at 100
$\micron$.

The resulting model SED is plotted in
Fig.~\ref{dusty:model}. The flux density at wavelengths longer than
100 $\micron$ comes mainly from the envelope. Since the envelope is
bigger than the aperture size, the modeled flux density is higher
than the observations. To compare to observations, we used OBSSPHere
to convolve the model with the aperture that was used for the
photometry. The convolved model data are marked with open circles in
Fig.~\ref{dusty:model}. The SED shows a good fit to the observed data
at 24 $\micron$, 70 $\micron$, 160 $\micron$, 350 $\micron$, and 850
$\micron$. However, the modeled SED shows a 24 $\--$ 30 $\micron$
turnover and shallow 15.2 $\micron$ CO$_2$ feature, which is not an
ideal fit to the IRS spectrum.

\subsection{Two-dimensional radiative transfer models}\label{radmc_sec}

The best fit result with the one-dimensional radiative transfer model
fails to fit the IRS spectrum at wavelength 24 $\--$ 30 $\micron$.
The flux density in the mid-infrared region mainly comes from the
disk emission. There are limitations to models of disks averaged
over all inclinations. For a more realistic explanation of the MIR
spectrum, we used the two dimensional axisymmetric Monte Carlo code,
RADMC \citep{2004A&A...417..159D}. The RADMC program solves the
radiative transfer equation in 2-D circumstellar dust
configurations. Once a dust temperature distribution is determined,
RAYTRACE is used to create smooth SEDs for various inclination
angles. As with the 1-D models, we use some constraints to determine
some parameters before running a fine grid; this reduces the
multi-dimensional fitting problem to a manageable size.

We used the same dust opacity and ISRF models used for
the 1-D model. For the 2-D model envelope, we used a slowly rotating
cloud core with a small angular momentum
(\citealt*{1984ApJ...286..529T}, hereafter TSC). The initial state of
the TSC model is a singular isothermal sphere, with density
proportional to $r^{-2}$. As the core evolves, the collapse of the
isothermal sphere includes the effects of an initially uniform and
slow rotation. The density of the TSC envelope is determined by the
initial angular velocity, $\Omega_0$ and the time, $t$. We set
$\Omega_0 = 10^{-14} s^{-1}$, which is the value used in TSC
model. As $t$ increases, the infall radius gets larger and the density
distribution gets flattened. The best fit density distribution is
constrained, as described for the 1-D model, to correspond to
$t=35000$ yr. We use the inner radius and the outer radius as 350 AU
and 35000 AU respectively, which were determined by the 1-D
model and the core size. The mass of the envelope that matches the
long-wavelength emission is 5.5 \msun, only slightly larger than  the
1-D model estimate in \S 5.1.

For the disk density structure, we use the axisymmetric flared disk
\citep{2001ApJ...560..957D, 2003ApJ...591.1049W,
 2007ApJ...656..991P}. The central disk structure is as follows:
\begin{equation} \rho (R,z)= \frac{\Sigma (R)}{H_p (R) \sqrt{2
\pi}} exp \left[-\frac{z^2}{2H_p(R)^2} \right],
\end{equation}
where the surface density $\Sigma$ is
\begin{equation}
\Sigma (R) = \Sigma_{disk} \left( \frac{R}{R_{disk}} \right)^{-p} ,
\end{equation}
and the disk scale height is given as
\begin{equation}
\frac{H_p (R)}{R} = \frac{H_{disk}}{R_{disk}} \left(
\frac{R}{R_{disk}} \right)^{\alpha_{fl}}.
\end{equation}

We set the disk outer radius as 350 AU, which is the inner radius of
the envelope. The disk inner radius is 0.03 AU, set to the location at
which a dust temperature is 1500 K.  We use most of the other disk model
parameters in \citet{2003ApJ...591.1049W}.
The flaring index $\alpha_{fl}$ is set
to 0.25, and the surface density power law index is set to $p = 1$.
The disk mass is set as M$_{disk}$=0.01 \msun.

\begin{figure}[b]
\includegraphics[width=0.9\columnwidth]{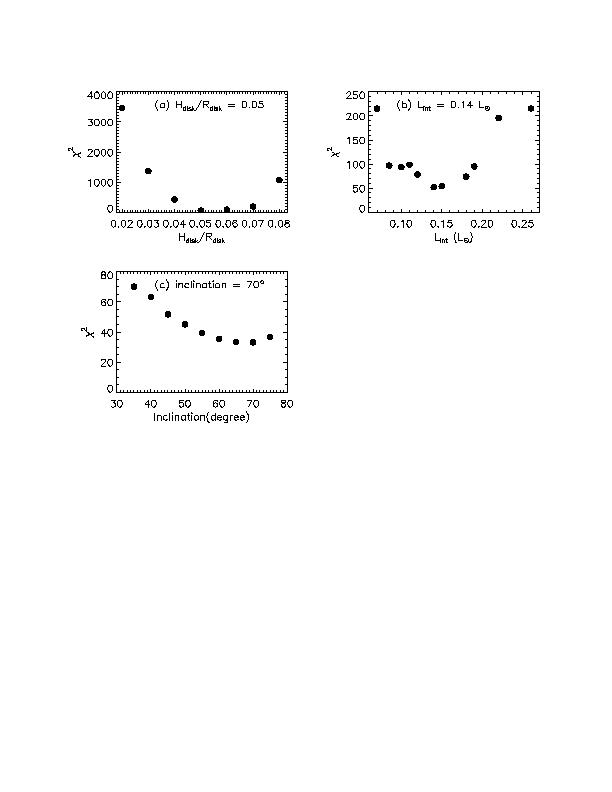}
\caption{\label{RADMC_chisq} The $\chi^2$ plots for a grid of
two-dimensional radiative transfer models. Panel (a) is
$\chi^2$ versus $H_{disk}/R_{disk}$, with $L_{int} = 0.14$ \lsun,
and inclination angle 45 degrees. The inclination angle 0 is
face-on. The smallest $\chi^2$ is at $H_{disk}/R_{disk} =0.05$.
Panel (b) is $\chi^2$ versus $L_{int}$, with $H_{disk}/R_{disk}
=0.05 $, and inclination angle 45 degrees. The smallest $\chi^2$ is
at $L_{int} = 0.14$ \lsun. Panel (c) is $\chi^2$ versus inclination
angle, with $H_{disk}/R_{disk} =0.05 $, and $L_{int} = 0.14$ \lsun.
The smallest $\chi^2$ is at inclination angle 70 degrees.
}
\end{figure}

\begin{figure}[t]
\includegraphics[width=0.9\columnwidth]{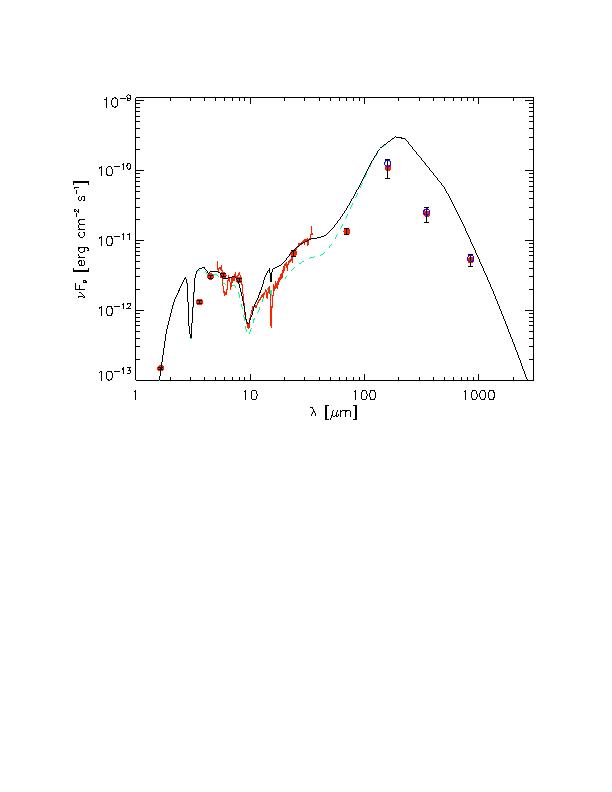}
\caption {\label{RADMC_sed}  SEDs obtained from the 2-D radiative
transfer modeling. The red filled circles are the photometry data
from the \textit{Spitzer}, CSO, and JCMT, and the red solid line is the
spectrum from the \textit{Spitzer} IRS. The black solid line is the
SED obtained from the model at inclination angle of 50\degree. The blue open
circles are flux densities at wavelengths where the source model has been
convolved with the  aperture used for measuring fluxes.
The green dashed line shows the model SED for an inclination angle
of 70\degree\ for comparison.
}
\end{figure}

We ran a grid of models varying three free parameters to fit the
observed SED. As for the 1-D models, we calculated $\chi^2$ for the
photometric points to determine the best fit.
The free parameters are the disk scale height, $H_{disk}/R_{disk}$ at
$R_{disk}$=100 AU, the internal luminosity, $L_{int}$, and the
inclination angle. Again, we plot the $\chi^2$ versus the
free parameter, setting the other two free parameters at particular
values. Panel (a) of Fig.~\ref{RADMC_chisq} presents $\chi^2$
varying the disk scale height $H_{disk}/R_{disk}$ in the range
between 0.01 \citep{2003ApJ...591.1049W} and 0.125
\citep{2007ApJ...656..991P}.  The best fit occurs
for $H_{disk}/R_{disk} = 0.05 \pm 0.005$. Panel (b) in
Fig.~\ref{RADMC_chisq} is the $\chi^2$ vs internal luminosity. Since
the disk is now heated only from the reprocessing of the central
star, the internal luminosity affects the SED mostly in the MIR. If
the internal luminosity is low, it cannot heat the disk enough. If
$L_{int}$ is high, the modeled SED becomes stronger than the
observed flux density everywhere.
The best fit is at $L_{int} = 0.14 \pm 0.005$ \lsun, which is a
little bit lower than found for the one-dimensional model (0.16 \lsun).
Panel (c) is the $\chi^2$ plot as a function of the inclination
angle.  The inclination angle affects the MIR SED, since the
inclination angle affects the disk emission.  There is a broad minimum in
$\chi^2$ around an inclination angle of 70\degree, but the SED for the
model at 70\degree\  badly underestimates the observations at 24 \micron,
which are
well fitted by a model with inclination angle of 50\degree, which however badly
overestimates the observations at 70 \micron. The
SEDs with these two inclination angles are plotted in
Fig.~\ref{RADMC_sed}. Neither model matches all of the IRS spectrum
either. In particular, the observed CO$_2$ feature is deeper than produced by
any of the models.

The best fit parameters are summarized in Table.~\ref{tabel:radmodel}.
The internal luminosity from the one-dimensional model is 0.16 \lsun,
the internal luminosity from the two-dimensional model is 0.14
\lsun. Both luminosities are slightly higher than the VeLLO
criterion ($L_{int} = 0.1$ \lsun). CB130-1-IRS1 is a low luminosity
source, but not a VeLLO.

\section{Molecular Line Model}\label{line_sec}

Chemical modeling of the line observations is needed to determine
the chemical properties of the source. We found that the chemical
properties can give further clues to the evolutionary state of the source.
In this section, we use step-function models of the chemical
abundances, which will provide a simple comparison to the full
chemical models employed in the next section.

Because the chemical models are not very sensitive to the 2-D structure
of the source, we use 1-D models of the dust. For each
chemical model, we first calculate the dust radiative transfer to get
the dust temperature distribution.  Next, we calculate the gas kinetic
temperature (\S~\ref{gastemp_sec}), and then use step function
abundance models, which can give a reasonable fit to the observed
molecular lines \citep{2003ApJ...583..789L, 2005ApJ...626..919E}
(\S~\ref{step_abun_sec}). A Monte Carlo code
\citep{1995ApJ...448..742C} calculates the molecular excitation and a
virtual telescope simulation produces line profiles, which can be
compared to the observed lines (\S~\ref{MC_simulation}). Comparing
observations to line profiles predicted from empirical abundance
profiles can constrain the abundance profiles.

\subsection{Gas Temperature}\label{gastemp_sec}

We calculate the radial gas temperature distribution using a gas
energetics code \citep{1997ApJ...489..122D, 2004ApJ...614..252Y},
which includes energy transfer between gas and dust, heating by cosmic
rays, photoelectric heating and molecular cooling.

The collision rate between gas and dust depends on the cumulative
grain cross-section. In this simulation, the grain cross section per
baryon is $6.09 \times 10^{-22}$ cm$^2$ \citep{2004ApJ...614..252Y},
based on the OH5 model. This is slightly different from the opacity
model we have used for DUSTY modeling, but the line modeling does not
critically depend on the opacity model. We take the cosmic ray
ionization rate as $3.0 \times 10^{-17}$ s$^{-1}$
\citep{2002A&A...395..233R, 2000A&A...358L..79V}.

The photoelectric heating is determined by the strength of UV part of
the ISRF and the electron number density. The electron number density
is assumed to be 0.001 cm$^{-3}$ \citep{2002A&A...389..446D}. The CO
line shows stronger emission when the envelope is heated by the
unattenuated ISRF. So we vary the extinction of the ISRF by the
surrounding material until it reproduces the observed CO line. See
\citet{2005ApJ...626..919E} for a full explanation of this method. The
relevant part of the ISRF for the gas heating is in the FUV, where the
ISRF is characterized by $G_0$. Assuming a plane parallel slab, the
ISRF is attenuated following $ G_0 = G(ISM) e^{-\tau_{UV}}$ where
$G(ISM)$ is the unattenuated FUV field. The best match  to the
observed CO line requires $G_0/G(ISM) = 4.5\ee{-3}$. The relation
between optical depth in the UV band and the V band is $\tau_{UV} =
1.8 A_V$. Thus we attenuate the ISRF at the outer boundary of the
envelope by $A_{V} =  -0.56 ln(4.5 \times 10^{-3}) = 3$. We used this
value for the external attenuation of the ISRF in \S
\ref{DUSTY_sec}. Of course, the FUV is further attenuated as it
propagates into the cloud, and the model accounts for this.

The cooling rate mainly depends on the abundance and the line width of
CO. We set the CO fractional abundance relative to H$_2$ to be $7.4
\times 10^{-5}$ \citep{2002A&A...389L...6B, 2002A&A...389..908J}. We
use a Doppler $b=0.8$ \kms, which fits best the CO molecular line.

\begin{figure}[t]
\includegraphics[width=0.8\columnwidth]{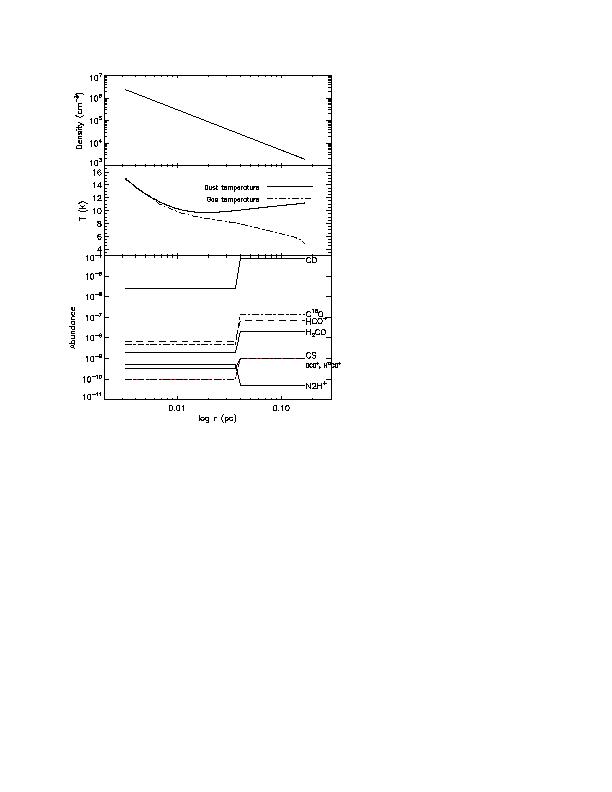}
\caption{\label{D_struct} The first panel is the density profile as
a function of radius, which follows $n(r) \propto r^{-1.8}$. The
second panel is the gas and dust temperature distribution as a
function of radius. The solid line is the dust temperature and the
dash-dot line is the gas temperature profile. The dust temperature
and the gas temperature are almost the same in the inner region due
to the high collision rate, because of the high density in the inner
part. In the outer part of the cloud the gas temperature is lower
than the dust temperature, since the photoelectric heating is too
low to heat the gas. The last panel is the step function abundance
profiles as a function of radius. The molecules freeze out in the
inner part of the cloud. Since CO destroys the N$_2$H$^+$ molecule,
the abundance profile of N$_2$H$^+$ has reversed shape to other molecules.}
\end{figure}

The calculated gas temperature is plotted in the middle panel of
Fig.~\ref{D_struct}, along with the dust temperature. The dust temperature
and gas temperature are almost the same in the inner region due to the
high gas-dust collision rate because of the high density in the inner
region, shown in the upper panel. In the outer part of the cloud the
gas temperature is lower than the dust temperature, because the gas
temperature has decoupled from the dust temperature, and the
photoelectric heating from the attenuated ISRF is too low to heat the
gas.

\subsection{Abundance Profile}\label{step_abun_sec}

We use step function abundance profiles.
Molecules tend to freeze out onto the grain
surfaces in cold and dense regions. The inner part of the cloud has
higher density compared to the outer region, so molecules freeze out
faster. If the dust temperature becomes high enough, the molecules
will sublimate. The CO sublimation temperature is about 20 K. The
highest temperature in the model cloud is less than 20 K in the
innermost region of the envelope, 0.003 pc, since CB130-1-IRS1 is a
low luminosity source. The dust temperature is too low to sublimate the
frozen-out CO. In the outermost part, the density is too low for
significant freeze-out. Thus, a step function is a reasonable
profile for the cloud. The three parameters of the abundance
profile are the undepleted abundance ($X_0$), depletion radius
($r_D$), and depletion fraction ($f_D$). The best-fitting parameters
are summarized in Table.~\ref{lab:abundance}. The abundance profiles
are plotted in the last panel in Fig.~\ref{D_struct}.

\subsection{Monte Carlo simulation}\label{MC_simulation}

The line profiles are modeled with the Monte Carlo (MC) code
\citep{1995ApJ...448..742C} to calculate level populations of
molecules. The MC code can handle arbitrary 1-D distributions of the
systematic velocity, the density, the kinetic temperature, the
microturbulence, and the abundance self-consistently.

Infalling envelopes sometimes exhibit a stronger blue wing for
optically thick lines with a substantial excitation temperature
gradient \citep{1999ARA&A..37..311E, 2007ApJ...671.1748L}. Even
though CB130-1 contains a protostar, there is no significant feature
of infall in molecular lines. All lines show a single peak, so we do
not include any systematic velocity in the models.

Doppler $b$ parameters are determined by the width of each line. We
have used $b=0.8$ \kms\ for CO. Since the CO line is optically thick
and dominated by the outer part of the envelope, the CO line is
difficult to fit with the same Doppler $b$ parameter as others. We
used $b=0.2$ \kms\ for all the other lines, based on the widths of
relatively optically thin lines, such as C$^{18}$O and H$^{13}$CO.

After MC modeling, a virtual telescope (VT) program is used to
integrate the emission along the line of sight and to match the
velocity and spatial resolutions, as well as the beam efficiency. All
lines are assumed to be centered at 7.77 \kms, determined by averaging
the LSR velocity of optically thin lines.

By comparing modeled spectra to observations, we can estimate the
abundance. The best fit abundances are given in
Table~\ref{lab:abundance}. We set the depletion radius as 0.04 pc.
We used the isotope ratio for the low-mass protostellar envelope
described in \citet{2004A&A...416..603J}. The ratio of O to $^{18}$O
is 540, and the ratio between C and $^{13}$C is 70. CO destroys
N$_2$H$^+$ molecules, so N$_2$H$^+$ has a reverse abundance profile
compared to the CO abundance. The N$_2$H$^+$ line has a hyperfine
structure. The MC code calculates line modeling for each hyperfine
line, and then they are added together.

The best fit results are shown in Fig.~\ref{Line:model}. The modeled
C$^{18}$O, CS, and N$_2$H$^+$ (\jj{1}0) lines show reasonable fits
to the observed lines, while modeled CO and HCO$^+$ lines show a poor fit
with the observed lines. The strengths are in a reasonable range for
both \hcop\ and H$^{13}$CO$^+$, but the abundance
that fits the observed H$^{13}$CO$^+$ line produces a self-reversed
feature in the  modeled HCO$^+$ line. The modeled HCO$^+$ has a low excitation
temperature in the outer region, so it shows absorption features at
the center of the line. However, the observed HCO$^+$ line does not
show a self-reversed structure. The modeled H$_2$CO also shows a self-reversed
structure. The prediction of self-reversed lines that are not observed
is a common problem \citep{1995ApJ...450..667C, 2009ApJ...705.1160C},
and it suggests that macroturbulent and clumpy 3-D models would be
more appropriate. Though modeled DCO$^+$, CS, and N$_2$H$^+$ (\jj{3}2)
line strengths are weaker than the observed lines, the difference is
less than a factor of 2.5.

\section{Chemo-dynamical Models}\label{chemical_evol_model}
\subsection{Chemical Models with Luminosity Evolution Scenarios}
CB130-1 is a low luminosity source. There is no significant
evidence of a CO outflow, but there is a reflection nebula
suggestive of an outflow cavity. CB130-1 could be
in the first hydrostatic core stage or in a
quiescent stage between episodic accretion bursts. With chemical
evolution modeling, we may be able to distinguish these
possibilities. Because it takes time to relax the chemistry in the
core, evidence of a past high luminosity stage might be present in
the abundances. We use the evolutionary chemo-dynamical model by
\citet{2004ApJ...617..360L} to test this idea.
This model calculates the chemical
evolution of a core from the prestellar core to the embedded
protostellar core stage. At each time step, the density profile, the
dust temperature, the gas temperature and abundances are calculated
self-consistently.

At the prestellar stage, the density distribution is described by a
sequence of Bonnor-Ebert spheres. We adopt the time duration at the
preprotostellar core stage in Table 4 in
\citet{2009ApJ...705.1160C}, which has a longer total timescale at
the prestellar stage. Once a singular isothermal sphere has been
reached with $n(r)\propto r^{-2}$, the inside-out collapse is
initiated, and $n(r)$ approaches $r^{-1.5}$ inside the infall
radius. The infall radius propagates outward at the sound speed
through the envelope.

We use three kinds of luminosity evolution models in the accretion
stage. First, in Model 1, we adopt the accretion luminosity from
\citet{2005ApJ...627..293Y} including the First Hydrostatic Core stage
(FHSC; \citealp{1969MNRAS.145..271L, 1993ApJ...410..157B,
1998ApJ...495..346M}). The FHSC represents the core in an initial
quasi-static contraction stage when it is still molecular. When the
core temperature reaches 2000 K, the molecular hydrogen dissociates,
so the core has a second collapse and the FHSC stage ends. Models
indicate that the FHSC has a short, but poorly determined, lifetime,
about 10$^4$ years, \citep{1995ApJ...439L..55B, 1998ApJ...495..346M},
a very low luminosity, and a high optical depth. While two candidates
have recently been proposed
\citep{2010ApJ...715.1344C,2010ApJ...722L..33E}, the FHSC
has not yet been clearly detected. In Model 1, the FHSC lasts 20000
years and the collapse of the FHSC occurs in 100 years.  When $t =
20000$ yr, $L_{acc} \sim 0.011$ \lsun, which is only 10\% of the
internal luminosity of CB130-1-IRS1. The luminosity increases rapidly
after 20000 years. We adopt  a time at the very end of the
FHSC stage of Model 1, which has a luminosity much lower than
that of CB130-1-IRS1.  To match the luminosity of the object,
we would have to be catching the
source in an extremely short-lived phase while the FHSC is collapsing.

The luminosity evolution of Model 2 is taken from \citet{2004ApJ...617..360L}.
In Model 2, the FHSC lasts 20000 years and the transition occurs in 32000
years, providing a better chance to see a source with the observed
luminosity. However, this longer transition stage has no particular
theoretical justification. Here, we choose 45000 years as our
desired time since $L_{acc}$ is 0.16 \lsun\ at that time.

\begin{figure}[t]
\includegraphics[width=0.9\columnwidth]{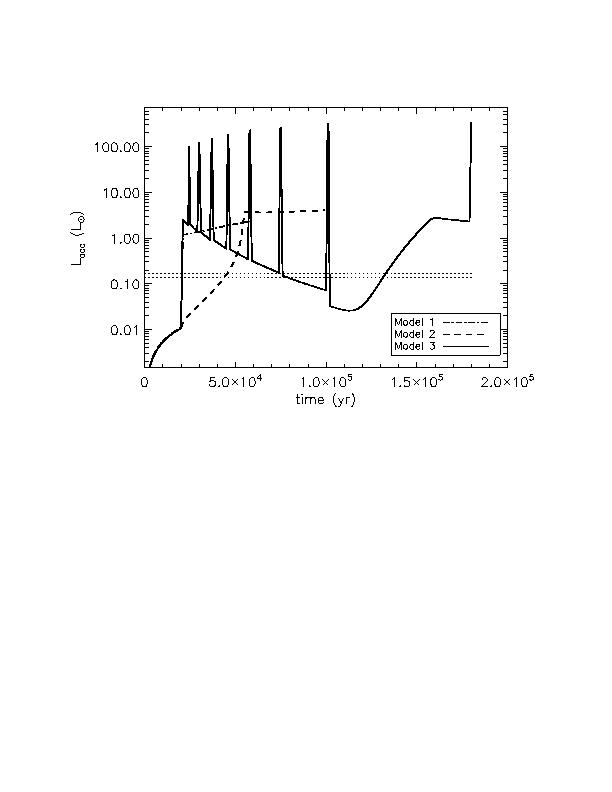}
\caption{\label{episodic_lum} The luminosity evolution of the models
we have used for chemical evolution. The dash-dot line is Model 1
\citep{2005ApJ...627..293Y}, the dashed line is Model 2
\citep{2004ApJ...617..360L}, and the solid line is Model 3
\citep{2010ApJ...710..470D}. The early stage evolution is the same
in all three models. The two horizontal lines indicate the best fit luminosities
for 1-D and 2-D models}
\end{figure}

The last model of luminosity evolution, Model 3,
uses the episodic accretion luminosity model
from \citet{2010ApJ...710..470D}. The episodic accretion
is included in a simple, idealized way rather than a self-consistent
model. The model assumes no accretion from disk to star most of the time,
but continued infall from envelope to disk.
A mass accretion burst occurs when the disk mass reaches
0.2 times the stellar mass. Then mass accretion from the disk to
protostar increases to $\dot{M} = 1 \times 10^{-4}$ \msun yr$^{-1}$,
which is about 100 times the average mass accretion rate. The
accretion luminosity depends on $\dot{M}$, so the accretion
luminosity also evolves episodically. The envelope mass of CB130-1
is 5.3 \msun $\--$ 5.5 \msun. So among the three masses in
\citet{2010ApJ...710..470D}, a 3 \msun\ envelope model is the
closest. However, the luminosity evolution of the model
with a 3 \msun\ envelope
never falls below 1 \lsun\ after the FHSC stage ends. So we
take the luminosity evolution of the 1 \msun\ envelope model in
\citet{2010ApJ...710..470D}. In Model 3, we adopt 78000 years as our
desired time step. At 78000 yr, the internal luminosity is 0.14
\lsun. The cloud has gone through 6 burst episodes, and
3000 years have passed after the sixth burst. The
luminosity evolution of the three models are plotted in
Fig.~\ref{episodic_lum}. The early stage evolution is the same in
all three models. After 20000 years, Model 2 has the lowest
luminosity among the three models.

We calculate the dust temperature distribution with DUSTY at each
time step. We use the same parameters described in
\S~\ref{DUSTY_sec}. Then we calculate the gas temperature
considering the parameters described in \S~\ref{gastemp_sec}. The
chemical evolution is calculated for each of 512 gas parcels as it
falls into the central region. Gas inside the infall radius carries
the memory of the conditions from farther out. The chemical
calculation includes interactions between gas and dust grains and
gas-phase reactions, but does not include chemical reactions on
grain mantles explicitly. We assume the surface binding energy of
species onto bare SiO$_2$ dust grains. We use an updated chemical
network with a new binding energy of N$_2$ and deuterated species,
which is used in \citet{2009ApJ...705.1160C}.

\begin{figure}[t]
\includegraphics[width=0.85\columnwidth]{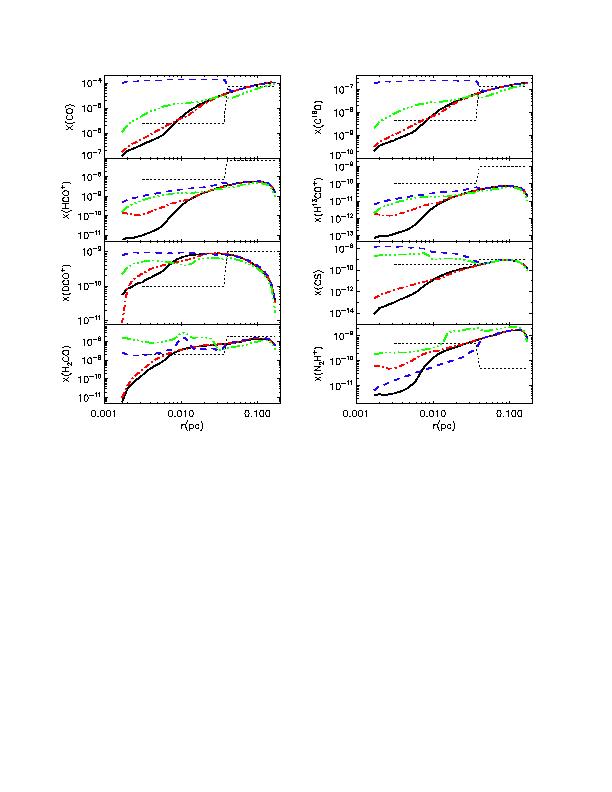}
\caption{ \label{abundance:chem} The abundance profiles with Model
1, Model 2, Model 3, Model 4, and the step function abundances for
comparison. The abundance profiles with Model 1 at 20000 years are
plotted in solid lines. The abundance profiles from Model 2 at 36000
years are plotted in dash-dot lines. The abundance profiles from
Model 3 at 78000 years are plotted in dashed lines. The
dash-dot-dot-dot lines are the abundance profiles of Model 4. For
comparison, we plotted the best fit step function abundances in
thin dotted lines. The chemical modeled abundance profiles are in the
range of the step function abundances except for \hcop\ and
H$^{13}$CO$^+$. }
\end{figure}

The abundance profile at each time step is calculated through the
chemical evolution model. We use the same isotope ratio as in
\S~\ref{MC_simulation}. The abundance profiles are plotted in
Fig.~\ref{abundance:chem}. The abundance profiles of Model 1 and Model
2 do not have a CO evaporation region at the center. In both cases,
the central source has a low luminosity, and it does not heat the
envelope enough to make CO sublimate. Model 2 has higher abundances
compared to Model 1, except for DCO$^{+}$. In Model 3, the abundances of
CO and C$^{18}$O show the CO sublimation region at the central
region because of past periods of high luminosity.
The modeled abundances of HCO$^+$, H$^{13}$CO$^+$ are all low
compared to the best fit step function abundances.

\begin{figure}[t]
\includegraphics[width=0.9\columnwidth]{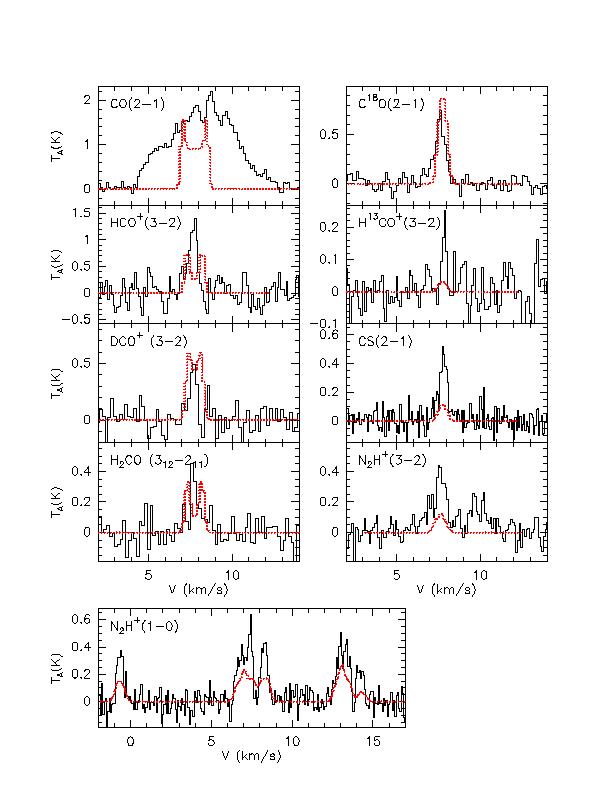}
\caption{\label{evol_0929_20000} The molecular lines from the
chemical evolution model with Model 1. The solid lines are the
observed molecular lines and the dotted lines are modeled molecular
lines. The time is 20000 yr with luminosity 0.011 \lsun, which is
just before the FHSC stage ends.}
\end{figure}

\begin{figure}[t]
\includegraphics[width=0.9\columnwidth]{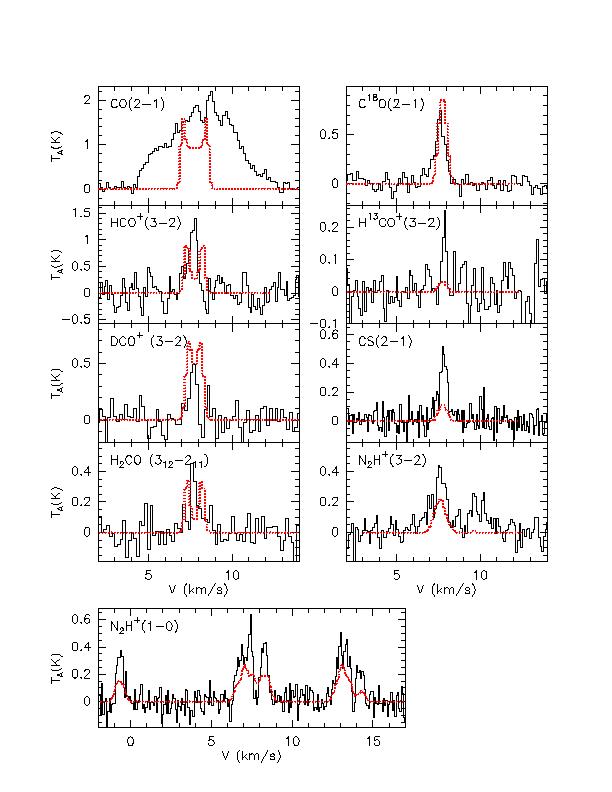}
\caption {\label{line_profile_FHSC} The molecular lines from Model
2. The solid lines are the observed molecular lines and the dotted
lines are modeled molecular lines. This model includes a long FHSC
transition stage. The
time is 45000 yr with the internal luminosity 0.16 \lsun. }
\end{figure}

\begin{figure}[t]
\includegraphics[width=0.9\columnwidth]{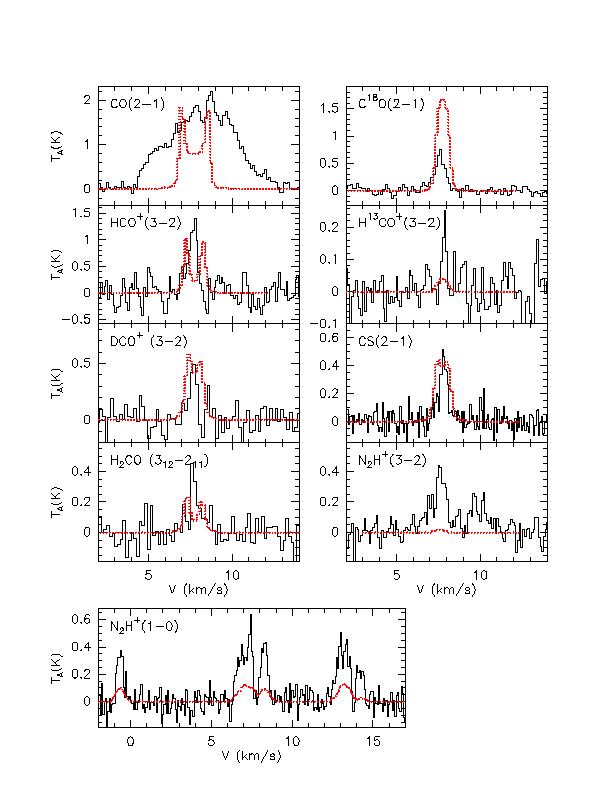}
\caption{\label{episodic_78000}  The molecular line result at 78000
years with Model 3. The solid lines are observed molecular lines,
and the dotted lines are the modeled lines at 78000 yr of Model 3.
The internal luminosity at this stage is $0.14$ \lsun. It has gone
through six periods of accretion bursts, and 3000 years have passed
after the accretion burst. The modeled C$^{18}$O is stronger and the
modeled N$_2$H$^+$ is weaker than the observed lines.}
\end{figure}

Once all the physical parameters are determined, we use MC code
\citep{1995ApJ...448..742C} to calculate the level populations. Then
we use VT code to produce line profiles. The molecular line profiles
with Model 1 are plotted in Fig.~\ref{evol_0929_20000}, and with Model
2 in Fig.~\ref{line_profile_FHSC}. The modeled lines of HCO$^+$,
H$^{13}$CO$^+$, CS, and N$_2$H$^+$ are weaker than the
observed lines, and the DCO$^+$ line is stronger than the observed
line. N$_2$H$^+$ shows a better fit with Model 2, but the \jj{3}2
line is still weak. The molecular lines with Model 3 are plotted in
Fig.~\ref{episodic_78000}. In Model 3, the C$^{18}$O line is much
stronger than observed lines. On the other hand, HCO$^+$,
H$^{13}$CO$^+$, and N$_2$H$^+$ are weaker than observed lines. The
modeled N$_2$H$^+$ (3-2) line is very weak. All three models
show strong C$^{18}$O and weak N$_2$H$^+$ lines compared to the
observed lines. The modeled C$^{18}$O is much
stronger in Model 3, because the episodic bursts have sublimated the
CO in the inner region.  The N$_2$H$^+$ line is weak in all three models.

All the models have caveats in their luminosity evolution
scenarios. Model 1 does not actually match the observed  L$_{int}$ $\sim$
0.14 \lsun $\--$ 0.16 \lsun.
Model 2 assumes that the transition stage of FHSC lasts
for 32000 years, which is unrealistic. It can give a source
luminosity of 0.16 \lsun, and it gives a slightly better fit to the
C$^{18}$O and N$_2$H$^+$ compared to Model 1. However, the modeled
\dcop\ line is too strong, and the modeled CS line is too weak. In Model 3, the
envelope mass is lower than our estimates for CB130-1. Also, the best fit
TSC model density from the 2-D model is 35000 years, while the
desired luminosity of Model 3 is reached at $t = 78000 $ years, which
means that more envelope material from Model 3 has already fallen in.
The problems with the luminosity evolution scenarios show how
chemical modeling can provide additional constraints that are distinct
from those from modeling the continuum.

\subsection{Including the Conversion of CO into CO$_2$ Ice}

\begin{figure}[t]
\includegraphics[width=0.9\columnwidth]{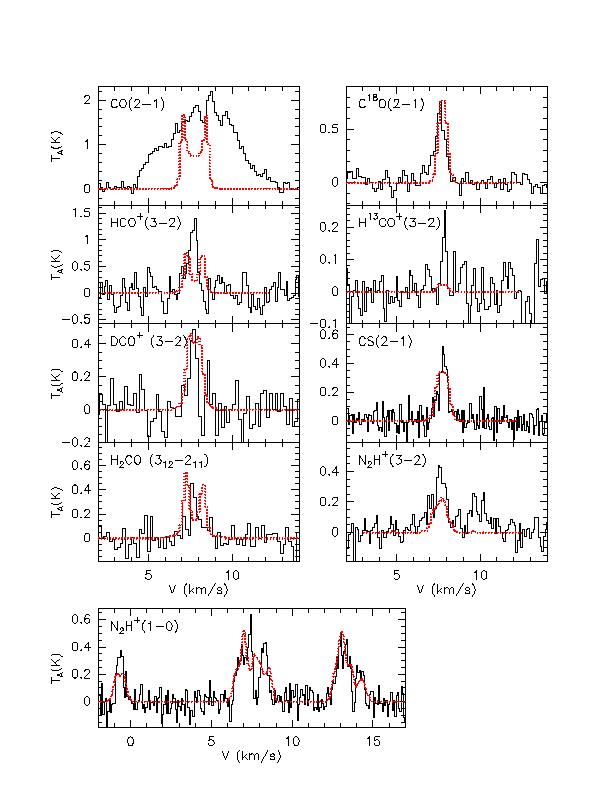}
\caption {\label{episodic_co80} The molecular line result with
Model 4. The solid lines are observed molecular lines, and the
dotted lines are the modeled lines. The evolution stage is the same
as Model 3. The calculation turned 80\% of CO ice into CO$_2$ ice
during low luminosity periods.  The lines of C$^{18}$O, CS, DCO$^+$,
H$_2$CO, and N$_2$H$^+$ are fitted better than in the other models.}
\end{figure}

We have one more trick to try. As noted in \S
\ref{result_CB130-1-IRS1}, a substantial amount of carbon is contained
in CO$_2$ ice. Once CO freezes out onto dust grain surfaces, a
fraction of CO ice turns into CO$_2$ ice
\citep{2007ApJ...656..980P}. CO$_2$ ice does not sublimate during the
accretion burst even when the CO evaporation temperature is
reached. So the amount of CO returned to the gas phase is less than
the amount of CO frozen onto grain surfaces during the quiescent
phase. The net effect is to decrease gas phase CO, and enhance
N$_2$H$^+$. We used the same episodic accretion luminosity evolution
model as Model 3, and varied the fraction of CO ice which turns into
CO$_2$ ice in the chemical network (Model 4). When 80\% of CO ice
turns into CO$_2$ ice, the modeled C$^{18}$O and N$_2$H$^+$ lines
match well with the observed lines. Also DCO$^+$ and CS fit the data
better than in Model 3. The abundance of this model is plotted in
Fig.~\ref{abundance:chem} and the molecular line profiles are plotted
in Fig.~\ref{episodic_co80}. The CO$_2$ ice column density obtained
from the best fit model is $2.94 \times 10^{18}$ cm$^{-2}$, 1.1 times
observed value, within the likely range of systematic uncertainties.

Among the models, the episodic model including CO$_2$ ice formation
fits best to the observed lines in all molecules. Based on the
modeling result, we suggest that CB130-1 has gone through episodic
accretion, forming CO$_2$ ice out of CO ice in quiescent phases. This
model can explain the low luminosity of the source, the strong CO$_2$
ice feature, and most of the gas phase emission lines. There are still
inconsistencies in this model, so further development of episodic
accretion models is needed.

\section{Summary}\label{summary}

We presented a detailed study of a low luminosity object in the CB130-1
region. We performed radiative transfer modeling and chemical
modeling to explain the core with a low luminosity protostar in it.

The embedded protostar CB130-1-IRS1 has $L_{int} = 0.14$ \lsun\
(1-D model) to $L_{int} = 0.16$ \lsun\ (2-D model). This is a
slightly higher luminosity than a VeLLO has, but still CB130-1-IRS1
is a low luminosity source at about 0.1 of the expected luminosity for
steady accretion onto an object at the boundary between stars and brown
dwarfs.

We tested both a step function abundance model and self-consistent
chemical evolution models. The step function abundance profile fits
observed lines reasonably well, but is of course, ad hoc. For the
self-consistent models, we use three different luminosity evolution
scenarios to explain the low luminosity of the object. All three
luminosity evolution models with the standard chemical network show
that the modeled C$^{18}$O line is strong and N$_2$H$^+$ is weak
compared to the observations. The step function model has more free
parameters than chemical evolution model, so it is more feasible to
fit the observed lines.

The deep 15.2 $\micron$ CO$_2$ ice feature indicates that CO$_2$ ice
has formed from CO ice. We added a reaction to the chemical network,
which turns CO ice into CO$_2$ ice. A model with that reaction decreases
the  C$^{18}$O abundance and increases the N$_2$H$^+$ abundance.
With the episodic accretion model and the modified network, we can
explain the low luminosity of the source, the strong CO$_2$ ice
feature, and most of the gas phase emission lines at the same time. So
CB130-1 is most likely neither a FHSC nor a very low mass object,
but a more evolved protostar in a quiescent stage between accretion
bursts. With a further study of CO$_2$, CO and N$_2$H$^+$ in low
luminosity objects, we may find chemical imprints of episodic
accretion in low luminosity objects.

We thank the Lorentz Center in Leiden for hosting several meetings
that contributed to this paper. Support for this work, part of the
\textit{Spitzer} Legacy Science Program, was provided by NASA
through contracts 1224608 and 1288664 issued by the Jet Propulsion
Laboratory, California Institute of Technology, under NASA contract
1407. Support was also provided by NASA Origins grant NNX07AJ72G and
NSF grant AST-0607793 to the University of Texas at Austin. This
research was also supported by the National Research Foundation of
Korea (NRF) grant funded by the Korea government (MEST) (No.
2009-0062866) and by Basic Science Research Program through the NRF
funded by the Ministry of Education, Science and Technology (No.
2010-0008704). TLB was partially supported by NASA through contracts
1279198, 1288806, and 1342425 issued by the Jet Propulsion Laboratory,
California Institute of Technology to the Smithsonian Astrophysical
Observatory, and by the NSF through grant AST-0708158.

%%%%%%%Begin Tables and Figures%%%%%%%
\begin{deluxetable}{rrrrrrrr}
\tablecolumns{8} \tablewidth{0pc}
\tablecaption{Photometry of CB130-1-IRS1}
\tablehead{
\colhead{$\lambda$} &
\colhead{$S_{\nu}(\lambda)$} &
\colhead{$\sigma$}    &
\colhead{Aperture diameter} &
\colhead{Telescope} \\
\colhead{($\micron$)} &
\colhead{$(mJy)$} &
\colhead{$(mJy)$} &
\colhead{ (arcsec)} &
\colhead{}
}
\startdata
1.25  & $<0.736$  & \nodata  & \nodata & CTIO \\
1.64  &  0.088    &  0.002   & 2       & CTIO \\
2.15  & $<1.27$   & \nodata  & \nodata & CTIO \\
3.6   &  1.58     &  0.08   & 1.7     & \textit{Spitzer}\\
4.5   &  4.57     &  0.23    & 1.7     & \textit{Spitzer}\\
5.8   &  6.12     &  0.31    & 1.9     & \textit{Spitzer}\\
8.0   &  7.28     &  0.36    & 2.0     & \textit{Spitzer}\\
24.0  &  52.4     &  5.2     & 6.0     & \textit{Spitzer}\\
70.0  &  312      &  32      & 50      & \textit{Spitzer}\\
160.0 &  5840     &  1700    & 100     & \textit{Spitzer}\\
350.0 &  2800     &  700     & 40      & CSO\\
850.0 &  1480     &  296     & 120     & JCMT\\

\enddata
\tablecomments{ The 350 $\micron$ continuum SHARC-II data is presented by
\citet{2007AJ....133.1560W}.}
\label{table-IRS1}
\end{deluxetable}

\begin{deluxetable}{rrrrrrrr}
\tablecolumns{8} \tablewidth{0pc}
\tablecaption{Photometry of CB130-1-IRS2}
\tablehead{
\colhead{$\lambda$} &
\colhead{$S_{\nu}(\lambda)$} &
\colhead{$\sigma$}    &
\colhead{Aperture diameter} &
\colhead{Telescope} \\

\colhead{($\micron$)} &
\colhead{(mJy)} &
\colhead{(mJy)}    &
\colhead{(arcsec)} &
\colhead{}

} \startdata
1.25 & 2.46 &  0.005  & 2  & CTIO \\
1.64 & 7.59 &  0.007  & 2  & CTIO \\
2.15 & 12.7 &  0.0    & 2  & CTIO \\
3.6  & 17.6 &  0.9    & 1.7& \textit{Spitzer}\\
4.5  & 18.2 &  0.9    & 1.7& \textit{Spitzer}\\
5.8  & 17.8 &  0.9    & 1.9& \textit{Spitzer}\\
8.0  & 20.5 &  1.0    & 2.0& \textit{Spitzer}\\
24.0 & 27.8 &  2.8    & 6.0& \textit{Spitzer}\\

\enddata
\end{deluxetable}

\begin{deluxetable}{lrrccccrr}
\tabletypesize{\scriptsize}
\setlength{\tabcolsep}{0.02in}
\tablecolumns{8}
\tablewidth{0pc}
\tablecaption{\label{line_obs_cb130-1} Observation parameters and
molecular line data for CB130-1}
\tablehead{
 \colhead{Line} &
 \colhead{Frequency} &
 \colhead{$\eta_{mb}$} &
 \colhead{Beam size} &
 \colhead{$\int T_A^* dv $} &
 \colhead{$v_{LSR}$}  &
 \colhead{$\Delta v$} &
 \colhead{$T_A^*$} &
 \colhead{Telescope} \\

 \colhead{} &
 \colhead{(GHz)} &
 \colhead{} &
 \colhead{(arcsec)} &
 \colhead{(K km s$^{-1}$)} &
 \colhead{(km s$^{-1}$)}  &
 \colhead{(km s$^{-1}$)} &
 \colhead{(K)} &
 \colhead{}

}
\startdata
CO 2-1          &230.537970& 0.7 & 33 & 9.38 &     8.27 &   4.73  &   1.86    & CSO \\
C$^{17}$O 2-1       &224.714368& 0.8 & 34 &\ldots&   \ldots & \ldots  & $ <0.14$  & CSO \\
C$^{18}$O 2-1       &219.560352& 0.8 & 35 & 0.48 &     7.61 &   0.71  &   0.64    & CSO \\
HCO$^+$ 3-2     &267.557620& 0.7 & 29 & 0.82 &     7.67 &   0.60  &   1.3     & CSO \\
DCO$^+$ 3-2     &216.112605& 0.7 & 36 & 0.26 &     7.68 &   0.47  &   0.53    & CSO \\
H$^{13}$CO$^+$ 3-2  &260.255339& 0.8 & 30 & 0.09 &     7.78 &   0.49  &   0.17    & CSO \\
N$_2$H$^+$ 3-2      &279.511701& 0.7 & 27 & 0.45 &     7.62 &   0.88  &   0.10    & CSO \\
N$_2$H$^+$ 1-0      &93.173700 & 0.6 & 70 & 2.27 &     7.35 &   1.07  &   0.99    & ARO \\
N$_2$H$^+$ 1-0      &93.176258 & 0.5 & 54 & 1.30 &     7.43 &   0.44  &   0.38    & FCRAO \\
CS 2-1          &97.980953 & 0.5 & 54 & 0.29 &     7.78 &   0.54  &   0.43    & FCRAO \\
N$_2$D$^+$ 3-2      &231.321635& 0.7 & 33 &\ldots&$\ldots$& $\ldots$  & $<0.24$   & CSO  \\
H$_2$D$^+$ 1,1,0 - 1,1,1&372.421340& 0.7 & 23 &\ldots&$\ldots$ & $\ldots$& $<0.29$ & CSO\\
H$_2$CO$3_{12} -2_{11}$ &225.697787& 0.7 & 34 & 0.21 &     7.74 &   0.53  &   0.37    & CSO  \\
\enddata
\end{deluxetable}

\begin{deluxetable}{lrccrrrrrr}
\setlength{\tabcolsep}{0.02in}
\tablecolumns{10}
\tablewidth{0pc}
\tablecaption{\label{line_obs_cb130-2}Observing parameters and molecular
line data for CB130-2.}
\tablehead{
 \colhead{Line} &
 \colhead{Frequency}&
 \colhead{$\eta_{mb}$} &
 \colhead{Beam size} &
 \colhead{$\int T_A^* dv $} &
 \colhead{$v_{LSR}$}  &
 \colhead{$\Delta v$} &
 \colhead{$T_A^*$} &
 \colhead{Telescope} \\

 \colhead{} &
 \colhead{(GHz)}&
 \colhead{} &
 \colhead{(arcsec)} &
 \colhead{(K km s$^{-1}$)} &
 \colhead{(km s$^{-1}$)}  &
 \colhead{(km s$^{-1}$)} &
 \colhead{(K)} &
 \colhead{}

}
\startdata
CO 2-1         &230.537970 & 0.7 & 33 & 10.1   &   7.99 &   5.54  &   1.72  & CSO \\
C$^{18}$O 2-1  &219.560352 & 0.8 & 35 & 0.40   &   7.19 &   0.56  &   1.04  & CSO \\
HCO$^+$ 3-2    &267.557620 & 0.7 & 29 & 0.23   &   7.23 &   0.47  &   0.47 & CSO \\
N$_2$H$^+$ 3-2 &279.511701 & 0.7 & 27 &\ldots  &$\ldots$& $\ldots$& $<0.13$ & CSO \\
N$_2$D$^+$ 3-2 &231.321635 & 0.7 & 33 &\ldots  &$\ldots$& $\ldots$& $<0.30$ & CSO \\
N$_2$H$^+$ 1-0 &93.176258  & 0.5 & 54 &\ldots  &$\ldots$& $\ldots$& $<0.24$ & FCRAO\\
CS 2-1         &97.980953  & 0.5 & 54 & 0.32   &   7.38 &   0.52  &   0.58  & FCRAO\\
\enddata
\end{deluxetable}

\begin{deluxetable}{lrccrrrrrr}
\setlength{\tabcolsep}{0.02in}
\tablecolumns{10}
\tablewidth{0pc}
\tablecaption{\label{line_obs_cb130-3} Observing parameters and
 molecular line data for CB130-3.}
\tablehead{
 \colhead{Line} &
 \colhead{Frequency}&
 \colhead{$\eta_{mb}$} &
 \colhead{Beam size} &
 \colhead{$\int T_A^* dv $} &
 \colhead{$v_{LSR}$}  &
 \colhead{$\Delta v$} &
 \colhead{$T_A^*$} &
 \colhead{Telescope} \\

 \colhead{} &
 \colhead{(GHz)}&
 \colhead{} &
 \colhead{(arcsec)} &
 \colhead{(K km s$^{-1}$)} &
 \colhead{(km s$^{-1}$)}  &
 \colhead{(km s$^{-1}$)} &
 \colhead{(K)} &
 \colhead{}
}
\startdata
CO 2-1         &230.537970 & 0.7 & 33 & 7.77 &   7.59 &   4.51  &   1.62  & CSO \\
C$^{18}$O 2-1  &219.560352 & 0.8 & 35 & 0.55 &   7.16 &   0.40  &   1.28  & CSO \\
HCO$^+$ 3-2    &267.557620 & 0.7 & 29 & 0.36 &   7.18 &   0.48  &   0.70  & CSO \\
N$_2$H$^+$ 3-2 &279.511701 & 0.7 & 27 &\ldots&$\ldots$& $\ldots$& $<0.13$ & CSO \\
N$_2$D$^+$ 3-2 &231.321635 & 0.7 & 33 &\ldots&$\ldots$& $\ldots$& $<0.15$ & CSO \\
N$_2$H$^+$ 1-0 &93.176258  & 0.5 & 54 & 0.59 &   7.90 &   0.23  &   0.41  & FCRAO\\
CS 2-1         &97.980953  & 0.5 & 54 & 0.42 &   7.39 &   0.83  &   0.54  & FCRAO\\
\enddata
\end{deluxetable}

\begin{deluxetable}{rrr}
\tablecolumns{3} \tablewidth{0pc}

\tablecaption{The best fit model parameters of CB130-1}
\tablehead{
\colhead{Parameter} &
\colhead{One-dimensional} &
\colhead{Two-dimensional}
}
\startdata
Stellar Temperature  &  3000 K                    & \ldots \\
Internal Luminosity  &  0.16 $\pm$ 0.005 \lsun & 0.14 $\pm$ 0.005 \lsun \\
Inner radius         &  350 AU                    & \ldots\\
Inclination angle    &  \ldots                    & 50-70\degree \\
Disk scale height    &  \ldots                    & 0.05 $\pm$ 0.005 \\

\enddata
\label{tabel:radmodel}
\end{deluxetable}

\begin{deluxetable}{rrrr}
\tablecolumns{4}
\tablewidth{0pc}
\tablecaption{ Best-fitting Step Function Abundance Parameters}
\tablehead{\colhead{Line} &
\colhead{$X_0$} & \colhead{$r_D ~(pc)$} & \colhead{$f_D$}} \startdata

CO             &  $7.40 \times 10^{-5}$  & 0.04 &  30 \\
C$^{18}$O      &  $1.37 \times 10^{-7}$  & 0.04 &  30 \\
HCO$^+$        &  $7.00 \times 10^{-8}$  & 0.04 &  10 \\
H$^{13}$CO$^+$ &  $1.00 \times 10^{-9}$  & 0.04 &  10 \\
DCO$^+$        &  $1.00 \times 10^{-9}$  & 0.04 &  10\\
N$_2$H$^+$     &  $5.00 \times 10^{-10}$ & 0.04 &  0.1\\
CS             &  $1.00 \times 10^{-9}$  & 0.04 &  3 \\
H$_2$CO        &  $2.00 \times 10^{-8}$  & 0.04 &  10 \\

\enddata
\tablecomments{The free parameter values of the step function
abundance profiles obtained with the line modeling.}
\label{lab:abundance}
\end{deluxetable}

\clearpage

\clearpage

\clearpage

\clearpage

\clearpage

\clearpage

\clearpage

\clearpage

\clearpage

\clearpage

\clearpage

\clearpage

\clearpage

\clearpage

\clearpage

\clearpage

\clearpage

\clearpage

\clearpage

\clearpage

\end{document}